\documentclass[arxiv,acm]{paper}

\begin{document}

\ifarxiv
\title{Building a Verifiable Logical Clock for P2P Networks}
\else
\title{\sys: A Verifiable Logical Clock for P2P Networks}
\fi

\ifauthor
\ifarxiv
\author{Guangda Sun}
\affiliation{
    \institution{National University of Singapore}
    \country{}
    \city{}
}
\email{sung@comp.nus.edu.sg}
\author{Tianyang Tao}
\affiliation{
    \institution{National University of Singapore}
    \country{}
    \city{}
}
\email{tianyangtao@u.nus.edu}
\author{Yanpei Guo}
\affiliation{
    \institution{National University of Singapore}
    \country{}
    \city{}
}
\email{guo_yanpei@u.nus.edu}
\author{Michael Yiqing Hu}
\affiliation{
    \institution{National University of Singapore}
    \country{}
    \city{}
}
\email{hmichael@nus.edu.sg}
\author{Jialin Li}
\affiliation{
    \institution{National University of Singapore}
    \country{}
    \city{}
}
\email{lijl@comp.nus.edu.sg}
\else
\author{Paper \#1008}
\fi
\else
\author{Working Draft}
\fi

\ifacm
\begin{abstract}

Logical clocks are a fundamental tool to establish causal ordering of events in a distributed system.
They have been applied in weakly consistent storage systems, causally ordered broadcast, distributed snapshots, deadlock detection, and distributed system debugging.
However, prior logical clock constructs fail to work in an open network with Byzantine participants.
In this work, we present \sys, a novel logical clock system that targets such challenging environment.
We first redefine causality properties among distributed processes under the Byzantine failure model.
To enforce these properties, \sys defines a new \emph{validator} abstraction for building fault-tolerant logical clocks.
Furthermore, our validator abstraction is customizable: \sys includes multiple backend implementations for the abstraction, each with different security-performance trade-offs.
We have applied \sys to build two decentralized applications, a mutual exclusive service and a weakly consistent key-value store.
\sys adds only marginal overhead compared to systems that tolerate no Byzantine faults.
It also out-performs state-of-the-art BFT total order protocols by significant margins.

\end{abstract} \fi

\maketitle

\ifacm
\else
\begin{abstract}

Logical clocks are a fundamental tool to establish causal ordering of events in a distributed system.
They have been applied in weakly consistent storage systems, causally ordered broadcast, distributed snapshots, deadlock detection, and distributed system debugging.
However, prior logical clock constructs fail to work in an open network with Byzantine participants.
In this work, we present \sys, a novel logical clock system that targets such challenging environment.
We first redefine causality properties among distributed processes under the Byzantine failure model.
To enforce these properties, \sys defines a new \emph{validator} abstraction for building fault-tolerant logical clocks.
Furthermore, our validator abstraction is customizable: \sys includes multiple backend implementations for the abstraction, each with different security-performance trade-offs.
We have applied \sys to build two decentralized applications, a mutual exclusive service and a weakly consistent key-value store.
\sys adds only marginal overhead compared to systems that tolerate no Byzantine faults.
It also out-performs state-of-the-art BFT total order protocols by significant margins.

\end{abstract} \fi

\section{Introduction}
\label{sec:intro}

The ordering of events is a fundamental concept in distributed systems.
In state machine replication systems~\cite{smr,paxos,vr,raft}, the set of replicas needs to agree on the order of operations in the log;
shards in a distributed database~\cite{spanner,megastore,calvin} are tasked to execute distributed transactions in a consistent partial order;
for mutual exclusion of shared resources, participants in a distributed system have to agree on the order of acquiring locks~\cite{chubby,lockmanager,netlock};
in a distributed storage system~\cite{gfs,ceph,bigtable,dynamo,pegasus}, servers apply a consistent order of mutations to storage objects.

It is well-known that perfectly synchronized clocks do not exist in realistic distributed systems, due to clock drift and relativity.
Ordering events using physical clock timestamps is therefore not reliable and can lead to anomalies.
Logical clocks~\cite{lamport-clock,vector-clock}, on the other hand, offer a solution to order events in a distributed system without relying on physical time.
Logical clocks order events based on \emph{causal relationship} instead of real-time information.
Concretely, if event $e_1$ potentially causes event $e_2$ to happen, a logical clock ensures the clock value of $e_1$ is smaller than the clock value of $e_2$.
Logical clocks have been applied to a wide range of distributed applications, including mutual exclusion~\cite{lamport-clock}, consistent distributed snapshot~\cite{snapshots}, eventual consistency~\cite{dynamo}, and causally consistent data stores~\cite{cops,bayou,tact}.

Many forms of logical clocks have been proposed in the literature~\cite{lamport-clock,vector-clock,bloom-clock,plausible-clock,tree-clock}.
Existing logical clock constructs, however, fall short in an open, decentralized network~\cite{bitcoin,ethereum}.
In these networks, any participant can join or leave the system at any time.
Such dynamic environment presents deep scalability challenges for vector-based logical clocks~\cite{vector-clock,plausible-clock,tree-clock}.
More critically, prior systems assume all participants in the system faithfully follow the protocol to update and propagate their clocks.
In a decentralized network, Byzantine~\cite{byzantine} behaviors, where a participant can deviate arbitrarily from the specified protocol, can make the clocks violate the causality properties, potentially corrupting application state or leading to undefined behavior.
Unfortunately, existing logical clock constructs are not Byzantine-fault tolerant.
Without reliable logical clocks, systems and applications are forced to use expensive BFT SMR protocols~\cite{pbft,hotstuff,bitcoin,honeybadger} which enforces total order of requests, even though their application semantics may only require partial ordering guarantees.

In this work, we propose \sys, a novel logical clock system that offers secure reasoning of causality for distributed applications in a P2P network.
At the core of \sys is a \emph{verifiable logical clock} (\clk) construct.
\clk provides similar ``happened-before'' partial order relation for events in a P2P network.
It uses a map structure to record clock values for each process, ensuring the size of each \clk is only proportional to the causal history and not to the system size.
A novel property of \clk is its \emph{verifiability}.
Each \clk is accompanied by a \emph{proof} that is verifiable by any third party.
By validating the proof, any process in the system can be convinced that the attached \clk is correctly generated by a \emph{sequence} of clock operations.
Moreover, the proof verifies that each clock input to the sequence of operations is also correct.
This \emph{recursive verifiability} is very powerful: It gives \clk the same causality correctness properties as previous logical clocks, even in the presence of Byzantine processes.
To improve usability of \clk, we further strengthen verifiability of each \clk to tolerate adversaries who intentionally create concurrent clocks, or generate erroneous clocks which deviates from application semantics.

We design \sys as a concrete instance of \clk.
\sys is a user-space library loaded into each distributed application.
To handle the complexity and diverse requirement of \clk verifiability, we define a \emph{validator} abstraction.
The abstraction is divided into a \emph{frontend} and a \emph{backend}.
The frontend defines the \emph{semantics} of each validator.
For \sys, we develop three such validator frontends:
An \emph{update validator} which verifies that the clock update operation obeys the \clk rules;
a \emph{monotonicity validator} to detect processes violating local ordering rules;
an \emph{application validator} that checks if the clock update follows application-specific semantics.
The backend, on the other hand, defines \emph{execution} of validators, \ie, how proofs are generated and validated.
We designed and implemented three such validator backends.
The \emph{quorum backend} relays the clock operation to a group a validator nodes.
Each validator checks the operation and signs a validation certificate.
The threshold signature from a group of validators serve as the proof.
The \emph{TEE backend} uses a validator with trusted execution environment (TEE) to verify the operation.
The backend uses attestation from the TEE as the proof.
Lastly, the \emph{IVC backend} uses a cryptographic construct called incrementally verifiable computation (IVC) to generate succinct recursive proofs without relaying on external validator nodes.

We leverage \sys to build two distributed applications which tolerate Byzantine behaviors.
The first use case develops a mutual exclusion protocol.
The protocol enforces the same set of guarantees as the original protocol~\cite{lamport-clock} in a fail-stop model, including ordered resource acquisition.
The second use case is a casually consistent key-value store, following the storage semantics of COPS~\cite{cops}.
The data store ensures client reads to obey causal consistency while allowing writes to be done without expensive global coordination.

We evaluate \sys and the use cases.
The end to end evaluations on the use cases show \sys secures verifiable causality with reasonable and sometimes even marginal overhead.
In the mutex evaluation, \sys adds less than 1\%/38\% latency to the unsafe baseline to handle one request, while a total order based approach would incur 4.3x latency to the baseline.
When all processes concurrently requesting, \sys is able to finalize all requests in 4.3s/18.8s with 60 processes.
In the data store evaluation, \sys is able to maintain over 200Kops/second throughput with ~100ms latency, while a total order replicated store can only reach 70Kops/second while latency already goes up to 1s.
In the micro-benchmarks, we study the various backends with fixed or varying computational resources.
Each of the backends is able to maintain steady proving time with up to 1000 clock size.
While circuit backend takes more than 1s to prove, the other backends are able to prove with millisecond scale latency.
The backends also scales well regarding clock size, the number of CPUs, the number of merged clocks and the number of faulty quorum members.

%
\section{Background and Related Work}
\label{sec:background}

\subsection{Causal Event Ordering}
\label{sec:background:causality}

Determining the order of events is a fundamental and important problem in distribute systems.
It is well-known that physical clocks are an unreliable source of event ordering due to relativity and clock skews.
The seminal work by Lamport~\cite{lamport-clock} addresses the issue by introducing a \emph{happened-before} relationship that defines the possible causality between events in a distributed system.
Specifically, let $\prec$ be a binary relation between pairs of events in an execution of a distributed system.
$e_1 \prec e_2$ if event $e_1$ may influence event $e_2$, or equivalently, $e_2$ is causally dependent on $e_1$.
$\prec$ is a strict partial order, \ie, it is irreflexive, asymmetric, and transitive.
Being a partial order, not all pairs of events are causally dependent.
If neither $e_1 \prec e_2$ nor $e_2 \prec e_1$, $e_1$ and $e_2$ are defined to be logically concurrent (represented as $e_1 \parallel e_2$).
Without perfectly synchronized physical clocks, it is impossible to determine which of $e_1$ or $e_2$ happens first if $e_1 \parallel e_2$.

Events in an execution are categorized into three general types.
Local events, \elocal, are any event happened on a process that does not involve messages;
\esend is a message send event from a source process, and for each \esend, there is a corresponding message receive event, \erecv, on the destination process if the message is successfully delivered.
Note that the model assumes all local events on a process happen \emph{sequentially}.

The happened-before relation $\prec$ on the set of events in an execution obeys the following rules:
\begin{itemize}
    \item \el{1} $\prec$ \el{2} if both events happen on the same process and \el{1} happens before \el{2} in the local sequential event order.
    \item \esend $\prec$ \erecv if \esend and \erecv are the corresponding message send and receive pair.
\end{itemize}

\subsection{Logical Clocks}
\label{sec:background:lc}

Logical clocks are a common approach to determine the happened-before relation defined in \autoref{sec:background:causality}.
One instance of logical clocks is the Lamport clock~\cite{lamport-clock}.
Using Lamport clock, each process $n_i$ in the system maintains a local clock $c_i$, represented as a natural number.
Upon a local event \el{i}, $n_i$ increments its local clock by one.
When $n_i$ sends a message, $n_i$ increments its local clock, and attaches the local clock value in the message.
When $n_i$ receives a message with a clock value $c_m$, it sets its local clock to $max(c_i, c_m) + 1$.

The logical time of an event $e$, represented as $c_e$, is the local clock value after the clock update.
Lamport clock guarantees the following property: If $e_1 \prec e_2$, then $c_{e_1} < c_{e_2}$.
However, the inverse is not true, \ie, $c_{e_1} < c_{e_2}$ does not imply that $e_1 \prec e_2$.
To be more precise, if $c_{e_1} < c_{e_2}$, either $e_1 \prec e_2$ or $e_1 || e_2$, but not $e_2 \prec e_1$.

Vector clock~\cite{vector-clock,holy_grail} addresses this shortcoming of Lamport clock.
As the name suggested, a vector clock, $v$, consists of a vector of natural numbers.
Cardinality of a vector clock equals the size of the system.
Each process is assigned a unique index in the vector clock.
We use $v[i]$ to denote the $i$th number in a vector clock $v$.
Upon a local event, $n_i$ increments $v_i[i]$ by one.
When $n_i$ sends a message, $n_i$ increments $v_i[i]$, and attaches $v_i$ in the message.
When $n_i$ receives a message with a clock $v_m$, it sets $v_i$ to $v_i'$, where $\forall p \in [0..S), v_i'[p] = max(v_i[p], v_m[p])$, $S$ is the size of the system.
$n_i$ then increments $v_i[i]$.

$v_i < v_j$ if and only if $\forall p \in [0..S), v_i[p] \le v_j[p]$ and $\exists p \in [0..S), v_i[p] < v_j[p]$.
By definition, there exists $v_i$ and $v_j$ such that neither $v_i < v_j$ nor $v_j < v_i$, \ie, $<$ is a partial order on the set of vector clocks.
Vector clock guarantees the following stronger property: $e_1 \prec e_2$ \emph{if and only if} $v_{e_1} < v_{e_2}$.

Logical clocks and causal event ordering have wide applicability to distributed systems and algorithms designs.
Lamport showed in his original paper how Lamport clocks can be used to implement distributed mutual exclusion~\cite{lamport-clock}.
Vector clocks are applied to realize consistent distributed snapshots~\cite{snapshots}.
Amazon leverages version vectors, a type of vector clocks, to build eventually consistent data store Dynamo~\cite{dynamo}.
Logical clocks are also used to implement various causally consistent key value stores~\cite{cops,bayou,tact}.

\subsection{Byzantine Fault Tolerance}

Failures are common events in a distributed system.
There are in general two types of failure models, crash failure model and Byzantine failure model.
In the crash failure model, once a process fails, it stops communication with all other processes until it recovers.
In the Byzantine failure model~\cite{byzantine}, a failed process can behave arbitrarily, including launching adversarial attacks to break the system.
Numerous distributed protocols have been developed to tolerate Byzantine faults.
The most common setting these protocols target is the state machine replication~\cite{smr} problem.
The core of the problem is reaching consensus on a \emph{totally ordered} request log, even in the presence of Byzantine behaviors.
The seminal work in this space is PBFT~\cite{pbft}, which is the first practical BFT protocol that tolerates up to $f$ Byzantine nodes using $3f + 1$ replicas.
PBFT is leader-based protocol.
After the leader orders and disseminates client requests, the backup replicas use two rounds of all-to-all communication to commit a batch of requests.
The protocol therefore incurs a latency of five message delays and an authenticator complexity of $O(N^2)$.
A long line of work subsequently improves various aspects of PBFT.
Zyzzyva~\cite{zyzzyva} uses speculative execution to reduce communication overhead.
It enables a fast path protocol that commits client requests in only three message delays.
HotStuff~\cite{hotstuff} adds one round of communication in normal operation to reduce the expensive view change protocol in PBFT from $O(N^3)$ to $O(N)$ authenticator complexity.
Several protocols~\cite{a2m,trinc,minbft} leverage trusted hardware, such as Trusted Platform Module (TPM)~\cite{tpm} and Trusted Execution Environment (TEE)~\cite{tee}, to reduce the replication factor to $2f + 1$.
A different line of BFT work~\cite{honeybadger,narwhal_tusk,dumbo} uses randomness to ensure protocol liveness even in a fully asynchronous network, bypassing the FLP impossibility result~\cite{flp}.

The work above all establish a total order of requests in the BFT setting.
Another recent line of work focuses on detecting causality and establishing causal ordering in the presence of Byzantine participants~\cite{bft_causality,bft-co,bft-co-tpds,bft-co-sac,bft-co-icpads,bft-co-nca,brb-co}.

\if false

This section covers background information on two main topics: causality of events in a distributed system, and verifiable computation.

\subsection{Verifiable Computation}

In systems where identities cannot be trusted (our target deployment model), publicly verifiable proofs are required to verify the claims made by the participants.
More concretely, if a node claims that the output of applying a certain function $\mu$ on input $x$ is $y$, the naive way to verify such a statement would be to re-execute the operation and compare the outputs.
Such an approach might not be viable when the verifier does not have enough computational resources to execute the function.
For instance, the current bitcoin blockchain is approximately 450 GB in size.
If a user wants to verify the latest block, he can either:

\begin{enumerate}
    \item Trust the person who provided the latest block to him (this is extremely unwise).
    \item Verify the whole chain himself from the genesis block; This takes a lot of compute time, storage space, and network bandwidth.
\end{enumerate}

An argument system is a cryptographic construct to achieve verifiable computation without trusting the entity performing the computation.
The goals of an argument system are quite simple.
For a given statement $\mu(x) \stackrel{?}{\rightarrow} y$, it produces an accompanying proof $\pi$.
This proof can be verified publicly to assert that the statement is true with all but a negligible probability.
More concretely, a prover $P$ wishes to convince a verifier $V$ that it knows some witness statement $w$ such that, for some public statement $x$ and arithmetic circuit $C$, $C(x,w)\rightarrow y$.

\paragraph{Properties of an argument system.}
There are two properties an argument system must satisfy:

\begin{enumerate}
    \item \textbf{Completeness:} A valid proof will always be accepted by a valid verifier.
    \item \textbf{(Knowledge) Soundness:} If a prover attempts to generate a proof without a valid witness, this proof will only be accepted by the verifier with a negligible probability.
\end{enumerate}

There are many types of argument system.
One of the most commonly used argument systems is Succinct Non-interactive Arguments of Knowledge (SNARK).
As the name suggested, using a SNARK, the verifier requires no further interaction with the prover, other than receiving the proof, when verifying;
the proof itself is also short, while the time to verify is fast (at most logarithmic to the circuit size).
If the witness $w$ cannot be derived by the verifier with sufficient probability, then the SNARK is also considered \emph{zero-knowledge} (zk-SNARK).
\sys does not require the zero-knowledge property, so we omit the details of zk-SNARKs.

\paragraph{Recursive proof systems.}

SNARKs are useful in many settings, e.g., cloud computing, as it allows a verifier to validate computationally expensive function executions in a fraction of the time to run it.
However, in some distributed computing scenarios, simply verifying a single execution is not sufficient.
Instead, we wish to verify a particular non-deterministic chain of executions.
Naively applying any off-the-shelf general circuit SNARK for every step will result in proofs and verification times that grows linearly.
In a highly evolving and volatile system, these metrics are unacceptable.

There has been some recent development in verifiable computation that has the potential to address the above challenges.
One particularly promising technique is recursive proof system.
In a recursive proof system, the prover recursively proves the correct execution of incremental computations.
Such technique can be applied to realize incrementally verifiable computation (IVC).
In IVC, in each step of the computation, the prover takes the output and proof of the previous step, and produces an output and proof for the next step.
A verifier only needs to verify the proof of a single step to ensure correct execution of the entire computation from genesis.
Critically, both prover and verifier time are independent of the length of the computation.
There exists quite a few constructions for recursive proofs in the wild, ranging from constructs like Halo\cite{Bowe2020RecursivePC} to PCD\cite{pcd}.
We envision more and more efficient constructs will be developed eventually.

A recursive proof system addresses the issue mentioned \autoref{sec:background:snarks} by having the proof be of a constant size regardless of the depth in the chain of executions. More concretely given a set of mutate functions $M=\{\mu_{1}, \cdots \},m_{n}=\{\mu_{n_1},\mu_{n_2}, \cdots \mu_{n_{end}}\}: m_{n} \subseteq M, \mu_{n_{end}}(\cdots \mu_{n_2}(\mu_{n_1}(s_0))) \rightarrow s_n$. Where $s_0$ is the genesis state, and $\forall i,j\in n, |\pi_i| = |\pi_j|$. The time-cost to verify should also approximately be the same.

For the sake of discussion, we imagine a \textit{black box} recursive proof system that contains at least the following two algorithms:

\begin{enumerate}
    \item \textbf{Prover: } $P(s_i,\pi_i,w_{i+1},d) \rightarrow (s_{i+1}, \pi_{i+1},d+1)$

    In this algorithm, the prover asserts that a valid operation has been applied on $op(s_i) \rightarrow s_{i+1}$. This operation can be congruent to $\mu \in M$, where $M$ is the family of valid functions. Furthermore, this proof also asserts that $s_i$ is a valid state: there exists an ordered set
    of mutate functions such that:
    \[
    M=\{\mu_{1}, \cdots \},m_{i}=\{\mu_{i_1},\mu_{i_2}, \cdots \mu_{i_{end}}\}: m_{x} \subseteq M, \mu_{i_{end}}(\cdots \mu_{i_2}(\mu_{i_1}(o_0))) \rightarrow o_i.
    \]
    Depth $d$ refers to the amount of functions that has been applied to the current state, from the original reference genesis state $s_0$.

    The witness might be non-deterministic. This means that there might exist infinite valid states that is immediately preceded by $s_i$.

    \item \textbf{Verifier: } $V(s_{i+1}, \pi_{i+1},d+1) \xrightarrow{?} \{1,0\}$.
    A correct verifier takes a state, and its corresponding proof to determine if the state is valid.
    If the proof is invalid, there is a negligible probability that the verifier asserts the state is valid. If the proof is valid, the verifier will always assert the state is valid.

\end{enumerate}

\subsubsection{General categories of Recursive Proof systems}
\mh{This section must be reviewed and updated after discussion with JiaHeng. }

There exists many proof systems or SNARKs, each with its unique characteristics. However, in the realm of recursive SNARKs, there exists mainly three categories. This section will briefly cover the differences. We will evaluate the utility of each recursive proof systems in autoref{sec:design:vlc}.

\begin{enumerate}
    \item \textbf{IVC: }
    Incrementally Verifiable Computation (IVC) by valiant is the ``father'' of recursive proof systems: Its creation led to the various recursive proof systems we have today.

    In the context of our system, given a genesis state $s_0$, the next state can be created by applying a function $\mu \in M, |M| = 1$. That is to say the $n^{th}$ state is the result of applying the function on $s_0$ $n$-times.
    At each state, a proof of constant size can be generated that asserts to its validity.

    IVC is applicable in \sys if a universal circuit is used; An universal circuit however might be quite expensive.

    \item \textbf{Non-Uniform IVC: }
    First introduced in SuperNova\cite{Kothapalli2022SuperNovaPU}. A non-uniform IVC is described as a generalization of IVC's in respect to the family of functions $M$.
    That is to say now $M = \{ \mu_1,\cdots,\}, |M| \geq 1$.

    Additionally, it introduces a new function $\varphi$ that takes in the inputs of the function $\mu_i(s_{i-1}) \rightarrow s_i$ and a potentially non-deterministic input to output the next function $\mu_{i+1}(s_i) \rightarrow s_{i+1}$.

    Intuitively, based on the existing state and the function $\varphi$, $\varphi$ creates a program by selecting a particular ordered multiset of functions from $M$.

    \item \textbf{PCD: }

    Proof carrying data (PCD), is described as a generalization of IVC's single chain structure to a Direct acyclic graph.

    In the context of our system, PCD will allow the creation of proofs for a given state $s_i$. The proof asserts that there exist a particular set of functions $m$ such that:
    \[M=\{\mu_{1}, \cdots \},m_{i}=\{\mu_{i_1},\mu_{i_2}, \cdots \mu_{i_{end}}\}: m_{x} \subseteq M, \mu_{i_{end}}(\cdots \mu_{i_2}(\mu_{i_1}(s_0))) \rightarrow s_i.
    \]

    Note that the set of functions $m$ is not necessarily hidden. A valid PCD proof just asserts that such a set exists.

\end{enumerate}

\begin{outline}

\textbf{Relationship Between Causality and State Mutation}

All revisions of a mutable state form a DAG, where the edges point from derived revision to base revisions, i.e., connect the inputs of mutation functions to its output.

Approach for tracking state mutation: tracking the causality. There's causality relationship between state revisions before and after the mutation. If we can track the causality among system events, then the subset of mutation events give us desired result.

\textbf{Tracking Mutable States in Data Center Networks}

Use the approach above, deploy techniques for tracking causality. Lamport clock and others.

Centralized managed.

\textbf{Tracking Mutable States in P2P Networks}

First line of works: global synchronization (consensus) on single proposer. PoW, PoS, etc. Poor scalability.

Second line of works: enable concurrent proposing but still linearizable. Peers must receive every message, even the ones they are not intereted in, to ensure there's no gap between the interesting messages.

Key: providing linearizable forces peers to process full history. Unnecessary for a system that allows branching, where only the branch the peer is working on is relevant.

\textbf{Recursively Verificable}

\sgd{background or clock design?}

Insight: define a clock for p2p networks, similar to the one used in data center approach, that brings partial ordering to mutable state.

Problem: have to verify every clock updating.

Solution: use IVC to finish verification in contant time.

\end{outline}

\fi
\section{A Verifiable Logical Clock}
\label{sec:clock}

We start the section with a motivating example to illustrate the impact of Byzantine behaviors on causal event ordering.
We then define a new type of logical clock, the verifiable logical clock (\clk), to address the issues.

\subsection{Motivating Example}
\label{sec:clock:example}

\begin{figure}
    \centering
    \includegraphics[width=0.8\columnwidth]{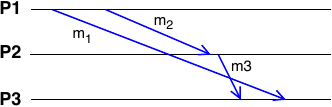}
    \caption{
        A motivating example.
    }
    \label{fig:example}
\end{figure}

We use the causal message delivery~\cite{reli_comm,causal_ordering} problem as a motivating example.
In a system that guarantees causal message delivery, if two messages $m$ and $m'$ are received by the same process, and $m$ is ordered before $m'$ by the happened-before relationship $\prec$ (\autoref{sec:background:causality}), then the process delivers $m$ before $m'$.
\autoref{fig:example} shows an example involving three processes, $P1$, $P2$, and $P3$.
The event of sending message $m_1$ happened before the message send event of $m_3$ ($\es{m_1} \prec \es{m_2}$, $\es{m_2} \prec \er{m_2}$, $\er{m_2} \prec \es{m_3}$ $\implies \es{m_1} \prec \es{m_3}$).
However, $P3$ receives $m_3$ before $m_1$.
This can happen even if processes use ordered, reliable point-to-point channels (\eg, TCP).
If the system enforces causal message delivery, $P3$ can only deliver $m_3$ after delivering $m_1$.

Vector clocks can be used to implement causal message delivery~\cite{causal_ordering}.
By following the protocol described in \autoref{sec:background:lc}, the vector clocks attached in messages $m_1$, $m_2$, and $m_3$ are $\text{\flqq} 1, 0, 0\text{\frqq}$, $\text{\flqq} 2, 0, 0\text{\frqq}$, and $\text{\flqq} 2, 2, 0\text{\frqq}$ respectively.
To ensure that all the received messages are delivered in causal order (but allows dropping of messages), $P3$ compares its local clock against the clock of every received message.
If the received clock is smaller (the partial order comparison as defined in \autoref{sec:background:lc}) than the local clock, the message is discarded.
When $P3$ first receives $m_3$ with clock $\text{\flqq} 2, 2, 0\text{\frqq}$, its local clock is smaller than the received clock, so $P3$ delivers $m_3$ and updates its local clock to $\text{\flqq} 2, 2, 1\text{\frqq}$.
When $P3$ later receives $m_1$, since the message clock $\text{\flqq} 1, 0, 0\text{\frqq}$ is smaller than $P3$'s local clock, $m_1$ is not delivered.

The original causal message delivery definition~\cite{reli_comm,causal_ordering}, however, does not allow rejection of message $m_1$.
To deliver $m_1$ while not violating causal ordering, $m_3$ needs to include extra information so that $P3$ is aware that it should wait for $m_1$ which happened before $m_3$.
Following the protocol proposed in \cite{causal_ordering}, each process maintains a vector clock mapped to every other process (called \textsc{ord\_buff\_s} in \cite{causal_ordering}).
Each vector clock maintains the maximum~\footnote{Note the maximum operation here is a per-element maximum, \ie, $v_{max} = {max}(v_a, v_b) \iff \forall p \in [0..S), v_{max}[p] = max(v_a[p], v_b[p])$} message clock \emph{destined} to the mapped process known by the local process.
The map is attached to each sent message.
In the example above, $P2$ updates the clock mapped to $P1$ to $\text{\flqq} 1, 0, 0\text{\frqq}$ after receiving $m_2$, and attaches this information in $m_3$.
When $P3$ receives $m_3$, it detects that its local clock $\text{\flqq} 0, 0, 0\text{\frqq}$ is smaller than the mapped clock value $\text{\flqq} 1, 0, 0\text{\frqq}$ in the message.
$P3$ then knows it has not received all messages with earlier causal dependencies, and therefore buffers $m_3$.
When $P3$ later receives $m_1$, the attached clock map is empty, so $P3$ delivers $m_1$ and updates local clock to $\text{\flqq} 1, 0, 1\text{\frqq}$.
Now the local clock is greater than $m_3$'s mapped value, so $P3$ can deliver $m_3$.

\subsection{Impact of Byzantine Processes}
\label{sec:clock:byzantine}

The above protocol assumes that all processes are honest and strictly follow the protocol.
Now, we show how Byzantine processes can violate the causal message delivery properties.
To simplify the discussion, we allow processes to reject messages that happened before previously delivered messages, so processes do not need to maintain and attach vector clock mapping \textsc{ord\_buff\_s}.
Note that since a Byzantine process can deliver messages in any arbitrary order, we can only enforce causal message delivery on non-faulty processes.
\lijl{We should probably also discuss omission failures.}

\paragraph{\textbf{Erroneous clock updates}}
A Byzantine process may violate the logical clock update rules as defined in \autoref{sec:background:lc}.
Suppose $P2$ is a Byzantine process in the scenario depicted in \autoref{fig:example}.
When $P2$ receives $m_2$ with vector clock $\text{\flqq} 2, 0, 0\text{\frqq}$, it could generate an erroneous vector clock $\text{\flqq} 0, 2, 3\text{\frqq}$ for message $m_3$.
As such, vector clocks for $m_1$ ($\text{\flqq} 1, 0, 0\text{\frqq}$) and $m_3$ ($\text{\flqq} 0, 2, 3\text{\frqq}$) are concurrent.
Subsequently, $P3$ is unable to detect the true causal dependency between $m_1$ and $m_3$, and will deliver $m_3$ before $m_1$.

\lijl{Should we also draw a figure for this case?}
Even a weaker form of clock forging can break causal message deliver.
Suppose a signature scheme is applied~\cite{signed_clock} so that a Byzantine process can only manipulate its own index in a vector clock.
Assume $P3$ has sent messages $m_a$ with clock $\text{\flqq} 0, 0, 1\text{\frqq}$ to $P2$ and $m_b$ with clock $\text{\flqq} 0, 0, 2\text{\frqq}$ to $P1$, before the message trace in \autoref{fig:example}.
$m_1$ and $m_2$ will now have vector clocks $\text{\flqq} 2, 0, 2\text{\frqq}$ and $\text{\flqq} 3, 0, 2\text{\frqq}$ respectively.
After $P2$ receives $m_2$, it now has two vector clocks locally, $\text{\flqq} 0, 0, 1\text{\frqq}$ (for $m_a$) and $\text{\flqq} 3, 0, 2\text{\frqq}$ (for $m_2$).
Even if $P2$ cannot forge clock values for other indices, it could still generate a vector clock $\text{\flqq} 3, 2, 1\text{\frqq}$ for $m_3$ by cherry-picking values from each received clock.
The clocks for $m_1$ ($\text{\flqq} 2, 0, 2\text{\frqq}$) and $m_3$ ($\text{\flqq} 3, 2, 1\text{\frqq}$) are concurrent;
$P3$ again will deliver $m_1$ after $m_3$.

\paragraph{\textbf{Local ordering violation}}
Lamport and vector clocks assume a sequential processor model, \ie, a process handles all events (including local events, message receive events, and message send events) sequentially.
An implication is that the logical clock of each process increases \emph{monotonically}: for any two events $e_1$ and $e_2$ happened on a process $p$, suppose $e_1$ occurred before $e_2$ in the local sequential order, then $e_1 \prec e_2$ and $c_{e_1} < c_{e_2}$.
To enforce this monotonicity property, a process maintains a single local clock which it sequentially updates on each event.

A Byzantine process may violate this local causal ordering property, even if it follows all clock update rules.
\lijl{Same question: should we draw another figure for this case?}
\lijl{This example is not entirely correct. Need to rethink.}
\sgd{The example you are seeking probably involves concurrent clocks generated by the same process.
This implies the merging pattern of events.
The only way to build such case with current message pattern in the figure is to assume a faulty P3, because the only merging pattern happens on P3.
P3 should first merge with receiving m3, then rewind its clock and merge again to m1.
Then it can confuse the other processes with the two concurrent clocks.}
Suppose the Byzantine process $P2$ generates a clock $\text{\flqq} 0, 3, 0\text{\frqq}$ and attaches it to a message that it sends to $P1$.
The same message pattern then occurs following \autoref{fig:example}.
$m_1$ and $m_2$ will have vector clocks $\text{\flqq} 2, 3, 0\text{\frqq}$ and $\text{\flqq} 3, 3, 0\text{\frqq}$ respectively.
When $P2$ receives $m_2$, instead of updating based on its current clock $\text{\flqq} 0, 3, 0\text{\frqq}$, $P2$ uses one of its older clocks, $\text{\flqq} 0, 1, 0\text{\frqq}$ for clock update.
Note that this is a legitimate clock generated by $P2$, so this is not an erroneous clock update.
$P2$ attaches the resulting clock, $\text{\flqq} 3, 2, 0\text{\frqq}$, to $m_3$.
Since $\text{\flqq} 3, 2, 0\text{\frqq}$ and $\text{\flqq} 2, 3, 0\text{\frqq}$ are concurrent, $P3$ delivers $m_1$ after $m_3$ which violates causal ordering.

\subsection{Verifiable Logical Clocks}
\label{sec:clock:properties}

We now define a new form of logical clock, a \emph{verifiable logical clock} (\clk), that tolerates the Byzantine behaviors described in \autoref{sec:clock:byzantine}.
\autoref{sec:design} elaborates on the concrete design of \clk.
We first define the system model, the clock operations, and the clock properties.
The system consists of a group of entities, each identified by a unique $id$.
We do not constraint the entities to be physical participants of the system;
they can also be virtual nodes, objects, or database keys.
We define two \clk operations:

\begin{itemize}
    \item $\fn{Init}() \rightarrow c$: Creates a genesis clock.
    \item $\fn{Update}(id, c, [c_1, c_2, \cdots]) \rightarrow c'$: Generates a new clock $c'$ by merging clock $c$ with a set of clocks $c_1, c_2, \cdots$ and then incrementing the merged clock.
    $id$ denotes the identifier of the entity that performs the increment.
    The set $\text{\flqq} c_1, c_2, \cdots\text{\frqq}$ could be empty.
\end{itemize}

Performing $\fn{Update}()$ with an empty set is equivalent to a local event clock update in prior logical clocks, while $\fn{Update}()$ with a single clock in the set is equivalent to a message receive event.
The set of \clk{}s forms a \emph{partially ordered set}.
The partial order relation $\prec$ on the \clk set is defined by the following rule: If $\fn{Update}(id, c, [c_1, c_2, \cdots]) \rightarrow c'$, then $c \prec c'$, $c_1 \prec c'$, $c_2 \prec c'$, $\cdots$

The partial order $\prec$ has the following properties:

\begin{itemize}
    \item \emph{Irreflexivity}: $\forall c \in C, c \nprec c$
    \item \emph{Asymmetry}: $\forall c_1, c_2 \in C, c_1 \prec c_2 \implies c_2 \nprec c_1$
    \item \emph{Transitivity}: $\forall c_1, c_2, c_3 \in C, c_1 \prec c_2 \land c_2 \prec c_3 \implies c_1 \prec c_3$
\end{itemize}

The above rules and properties also \emph{completely} define $\prec$, \ie, $c_1 \prec c_2$ \emph{if and only if} there exists a sequence of $\fn{Update()}$ operations that takes $c_1$ as an initial input and generates $c_2$.
An implication is that there exists \clk{}s that are not ordered by $\prec$.
For instance, given a genesis clock $c$ and two identifiers $id_1$ and $id_2$ such that $id_1 \neq id_2$, $c_1 \leftarrow \fn{Update}(c, id_1, [])$ and $c_2 \leftarrow \fn{Update}(c, id_2, [])$, then $c_1 \nprec c_2$ and $c_2 \nprec c_1$.
Two \clk{}s that are not ordered by $\prec$ are called \emph{concurrent}, and we use $c_1 \parallel c_2$ to denote concurrency.

Given the above definitions, we also define a compare function to determine the partial order of two \clk{}s:

\begin{itemize}
    \item $\fn{Compare}(c_1, c_2) \rightarrow \lbrace BF, EQ, AF, CC\rbrace$: Returns $BF$ if $c_1 \prec c_2$, $AF$ if $c_2 \prec c_1$, $EQ$ if $c_1 = c_2$, and $CC$ if $c_1 \parallel c_2$.
\end{itemize}

\paragraph{\textbf{Clock Verifiablility.}}
As discussed in \autoref{sec:clock:byzantine}, Byzantine processes can tamper with logical clock updates to violate causal order relationship among events.
To tolerate such Byzantine behaviors, \clk additionally provides \emph{verifiability}.
Each \clk $c$ is a tuple of $\tuple{v, \pi}$, where $v$ is the clock value and $\pi$ is a \emph{proof}.
The proof proves the following properties of \clk $c$ in the tuple:
\begin{enumerate}
    \item The clock is \emph{correctly} generated by a sequence of $\fn{Init}()$ and $\fn{Update}()$ operations.
    \item Each $\fn{Update}(id, c, [c_1, c_2, \cdots])$ in the sequence is invoked by a process with permission on the entity $id$.
    \item For each $\fn{Update}(id, c, [c_1, c_2, \cdots])$ in the sequence, each input clock ($c, c_1, c_2, \cdots$) \emph{recursively} satisfies the above properties.
\end{enumerate}

The proof can \emph{optionally} enforce the following global property: For each $id$ in the system, the set of \clk{}s correctly generated by $\fn{Update}(id, \dots)$ is \emph{totally ordered} by $\prec$.
Equivalently, no two clocks in the set are concurrent.

We only define the above property as optional since it is not required by all use cases.
We elaborate the impact of this optional property in \autoref{sec:design} and example applications that do not require it in \autoref{sec:cases}.

The system defines a $\fn{Verify}()$ function to validate the correctness of a \clk:

\begin{itemize}
    \item $\fn{Verify}(c) \rightarrow \lbrace\var{true}, \var{false}\rbrace$: Returns \emph{true} if the proof $\pi$ in $c$ is valid, and \emph{false} otherwise.
\end{itemize}

Note that invoking $\fn{Update}()$ and $\fn{Compare()}$ with invalid clocks will simply return an error.
Properly invoked $\fn{Init}()$ and $\fn{Update}()$ operations return clocks that are verifiable.
More formally:

\begin{itemize}
    \item If $\fn{Init}() \rightarrow c$, then $\fn{Verify}(c) \rightarrow \var{true}$
    \item If $\fn{Verify}(c) \rightarrow \var{true}$, $\fn{Verify}(c_1) \rightarrow \var{true}$, $\fn{Verify}(c_2) \rightarrow \var{true}$, $\cdots$, and $\fn{Update}(id, c, [c_1, c_2, \cdots]) \rightarrow c'$, then \\ $\fn{Verify}(c') \rightarrow \var{true}$
\end{itemize}

If the optional property of \clk proof is enforced, we need to augment the second clock verifiability rule by adding the clause: $c'$ is not $\parallel$ to any clock $c_{id}$ where $c_{id} \leftarrow \fn{Update}(id, \dots)$ and $\fn{Verify}(c_{id}) \rightarrow \var{true}$.

Similar to $\prec$, the above rules also \emph{completely} define \clk verifiability, \ie, $\fn{Verify}(c)$ returns $\var{true}$ \emph{if and only if} $c$ is generated by a sequence of properly executed $\fn{Init}()$ and $\fn{Update}()$ operations.

\paragraph{\textbf{Byzantine Fault Tolerance}}

Clock verifiability eliminates the causal message delivery violations in \autoref{sec:clock:byzantine}.
\lijl{Once we have figures for each of these violations, can easily refer to the figure.}
Suppose the ID for $P1$, $P2$, and $P3$ are $id_1$, $id_2$, and $id_3$ respectively.
Each process starts with an initial clock $c_0$ generated from $\fn{Init}()$.
In the first scenario, $P1$ generates \clk for $m_1$ and $m_2$ by $c_1 \leftarrow \fn{Update}(id_1, c_0, [])$ and $c_2 \leftarrow \fn{Update}(id_1, c_1, [])$.
By definition, $c_1 \prec c_2$.
When $P2$ receives $m_2$, the only \emph{valid} clock it can generate for $m_3$ is $c_3 \leftarrow \fn{Update}(id_2, c_0, [c_2])$\footnote{$P2$ can also use valid clocks other than $c_0$ as input to generate $c_3$. Regardless, the causal relationship between $c_3$ and $c_1$ will not change.}.
$P2$ does not have permission to other IDs when invoking $\fn{Update}()$, and $c_2$ is the only clock it receives from other processes.
$P3$ simply ignores any invalid clocks that do not pass $\fn{Verify}()$.
As such, $c_2 \prec c_3$ and therefore $c_1 \prec c_3$ by our clock properties.
$P3$ will thus reject $m_1$ after delivering $m_3$.

In the second weak forging scenario, clocks for $m_a$ and $m_b$ are generated by $P3$ as $c_a \leftarrow \fn{Update}(id_3, c_0, [])$ and $c_b \leftarrow \fn{Update}(id_3, c_a, [])$ respectively.
$P1$ will then generate $c_1 \leftarrow \fn{Update}(id_1, c_0, [c_b])$ for $m_1$, and $c_2 \leftarrow \fn{Update}(id_1, c_1, [])$.
As such, $c_a \prec c_b \prec c_1 \prec c_2$.
After $P2$ receives $m_2$, it possesses two valid clocks from other processes, $c_a$ and $c_2$.
\lijl{We do need to discuss omission issues earlier.}
When generating a valid clock $c_3$ for $m_3$, $P2$ can either include or exclude $c_a$ in the $\fn{Update}()$ invocation.
In either case, the resulting valid clock $c_3$ will have $c_2 \prec c_3$.

In the last local ordering scenario, suppose $P2$ generates a sequence of clocks, $c_a$, $c_b$, and $c_c$, by repeatedly invoking $\fn{Update}()$.
Therefore, $c_0 \prec c_a \prec c_b \prec c_c$.
It attaches $c_c$ to the first message it sends to $P1$.
When $P1$ receives the message and generates $c_1$ for $m_1$ and $c_2$ for $m_2$, we will have $c_c \prec c_1 \prec c_2$.
After $P2$ receives $m_2$, it can no longer use $c_0$, $c_a$, or $c_b$ as input to $\fn{Update}(id_2, c_{input}, [\cdots])$ and produce a valid clock, since any resulting clock $c_{output}$ will have $c_{output} \parallel c_c$.
This is a violation of the augmented clock verifiability rule.
\lijl{Complete this after correcting the example in \autoref{sec:clock:byzantine}.}

\lijl{We should move the following clock sealing part to design, together with discussions on application semantics.}

\section{\sys System Design}
\label{sec:design}

In \autoref{sec:clock}, we define the high-level properties of our new logical clock \clk.
In this section, we elaborate on the design of \clk which leverages a general \emph{validator} abstraction.
We then describe multiple concrete implementation of the validator abstraction that use quorum certificates, trusted hardware, and verifiable computation, each with its own trade-offs.

\subsection{Design Overview}
\label{sec:design:overview}

\begin{figure}
    \centering
    \includegraphics[width=0.9\columnwidth]{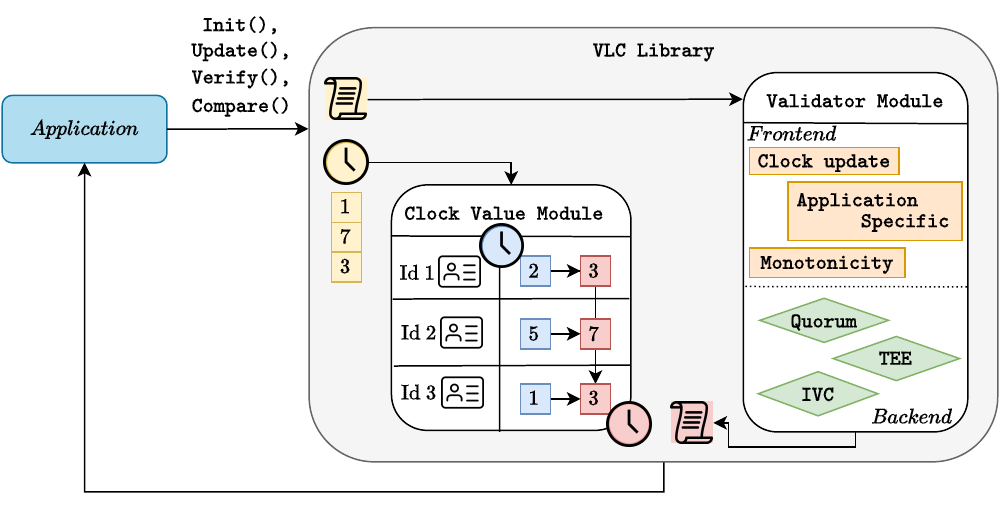}
    \caption{
        \sys overview. \lijl{Add more descriptions in the caption.}
    }
    \label{fig:overview}
\end{figure}

As shown in \autoref{fig:overview}, \sys is a user-space library linked into distributed applications.
The library exposes an API that includes the $\fn{Init}()$, $\fn{Update}()$, $\fn{Verify}()$, and $\fn{Compare}()$ functions as introduced in \autoref{sec:clock}.
The main data structure exposed to the application is an opaque \clk.
Internally, each \clk contains both numerical clock values and a proof (corresponding to the $\tuple{v, \pi}$ tuple in \autoref{sec:clock:properties}).
Besides using as input for the API functions, \clk also provides $\fn{serialize}()$ and $\fn{deserialize}()$ interfaces for the application to send/receive \clk{}s over the network.
\lijl{The API should include signatures and public keys.
There should also be a global configuration service that keeps ID, public key, and application-specific permission control rules.
Maybe describe them in the validator part.}

The \sys library internally consists of two main components: a clock module and a validator module.
The clock module defines the clock value structure in \clk (\ie, $v$ in $\tuple{v, \pi}$), and implements the basic clock update and comparison functionalities.
For each \clk, the module maintains a \emph{map} between an ID and an integer which presents the current clock value.
IDs not inserted in a \clk are implicitly mapped to clock value 0; a clock generated from $\fn{Init()}$ contains an empty map.
We use a map instead of a vector to support open systems with potentially unbounded participant set.
Size of each \clk is only linear to its causally dependent IDs.

When an application invokes $\fn{Update}(id, c, [c_1, c_2, \cdots])$, the module creates a new clock $c'$, and installs the \emph{maximum} clock value mapped to each present ID among $c, c_1, c_2, \cdots$ into $c'$.
It then increments the clock value mapped to the input $id$ in $c'$ by one.
To compare two clocks $c_1, c_2$ using $\fn{Compare}(c_1, c_2)$, the module returns $BF$ if the mapped clock value for each ID is not bigger in $c_1$ than that in $c_2$, and at least one of the values is smaller.
For the symmetrical opposite case, the modules returns $AF$.
It returns $EQ$ when all mapped values are equal.
For all other cases, the module returns $CC$.
Both the update and the comparison operation are similar to those in vector clocks.

The validator module is responsible for proving and verifying the \emph{validity} of \clk{}s, as defined in \autoref{sec:clock:properties}.
The module defines a general \emph{validator abstraction} (\autoref{sec:design:validator}).
The abstraction exposes a $\fn{prove}()$ and $\fn{validate}()$ interface to the library.
The $\fn{prove}()$ function includes instance-specific logic to test clock update invocations and generates a small \emph{proof};
$\fn{check}()$ verifies the validity of an input proof generated by $\fn{prove}()$.
\sys includes three concrete instances of the abstraction.
The \emph{update validator} checks the validity of a clock $\fn{Update}()$ invocation, as defined by the base clock verifiability properties in \autoref{sec:clock:properties}.
The \emph{monotonicity validator} ensures that the clock for each ID is increasing in a monotonic fashion, which corresponds to the optional clock verifiability property in \autoref{sec:clock:properties}.
The \emph{application validator} includes additional application-specific logic to further constrain $\fn{Update}()$ invocations;
such constraints were not covered in \autoref{sec:clock:properties}, and we elaborate their use cases in \autoref{sec:design:validator}.
Proofs generated by each validator instance are \emph{typed}.
We use $\pi_{up}$, $\pi_{mono}$, and $\pi_{app}$ and represent each proof type.
The proof in each \clk ($\pi$ in $\tuple{v, \pi}$) is effectively a union of $\pi_{up}$, $\pi_{mono}$, and $\pi_{app}$.
\lijl{Not all applications require all three proof types.}

We are now ready to give an overview of how the \sys library uses the two modules to process user API calls.
When $\fn{Verify}(c)$ is invoked, where $c$ is $\tuple{v, \tuple{\pi_{up}, \pi_{mono}, \pi_{app}}}$, the library inputs each of the $\pi_{update}$, $\pi_{mono}$, and $\pi_{app}$ to the $\fn{check}()$ functions of the corresponding validator instance.
The library returns $\var{true}$ if and only if all three $\fn{check}()$ calls return $\var{true}$.

When $\fn{Update}(id, c, [c_1, c_2, \cdots])$ is invoked, the \sys library first calls $\fn{Verify}()$ on each of the input clocks $c, c_1, c_2, \cdots$.
If all input clocks are properly verified, the library invokes the $\fn{prove}()$ function in the update, the monotonicity, and the application validator instances to get $\pi' \leftarrow \tuple{\pi_{up}, \pi_{mono}, \pi_{app}}$.
It then applies the update function in the clock module with $id$ and the input clock values to get $v'$.
The library then returns $c' \leftarrow \tuple{v', \pi'}$ to the application.

\lijl{Add $\fn{Init}()$ and $\fn{Compare}()$ overview.}

\subsection{The Validator Abstraction}
\label{sec:design:validator}

The critical property differentiating \clk from prior logical clocks is its \emph{verifiability}.
Traditional approaches such as public key signature schemes or verifiable computation~\cite{pantry,pinocchio,zaatar} establish verifiability for a single computational result.
\clk, however, requires \emph{recursive verifiability}: Validity of a \clk $c$ depends on not only the correct execution of the $\fn{Update}()$ function that generates $c$, but also \emph{recursively} on the validity of each of the input clocks.

Moreover, the exact semantics of \emph{logical clock correctness} depends on the application.
As we discussed in \autoref{sec:clock:properties}, the causal order relationship $\prec$ only requires the base clock verifiability properties.
For applications that additionally require \emph{local ordering} guarantees, \clk needs to enforce the augmented verifiability rule.
An application may also have specific restrictions on the set of logical clocks that can be generated.
As a concrete example, consider a storage application where clients include logical clocks as dependency set~\cite{cops, dynamo} in their \textsc{put} requests.
The application may require access control on the set of keys each client can write.
When generating logical clocks for new key versions, the clock update function needs to reject the request if the client lacks permission.
Such application-specific constraints are not enforced by the verifiability rules in \autoref{sec:clock:properties}.
We will elaborate this exact use case in \autoref{sec:cases:cstore}.

In \sys, we address these challenges by applying a \emph{modular} architecture with a common \emph{validator abstraction} for clock verifiability.
We forgo a monolithic design due to its inherit design complexity, and more importantly, its lack of flexibility.
Our validator abstraction defines a simple $\fn{prove}()$ and $\fn{check}()$ interface:
\begin{itemize}
    \item $\fn{prove}(\var{req}, \var{aux}) \rightarrow \lbrace \pi, \bot \rbrace$:
    Verify the validity of the $\fn{Update}()$ invocation in $\var{req}$.
    $\var{aux}$ is an optional auxiliary input for proof generation.
    Returns a proof $\pi$ if the request is valid, or $\bot$ otherwise.

    \item $\fn{check}(\pi) \rightarrow \lbrace{\var{true}, \var{false}} \rbrace$:
    Returns $\var{true}$ if the input proof $\pi$ is correct, $\var{false}$ otherwise.
\end{itemize}

We design three concrete instances of the validator abstraction.
Each instance applies specific clock verifiability rules:
the \emph{clock update validator} enforces the base properties defined in \autoref{sec:clock:properties};
the \emph{monotonicity validator} checks the optional local ordering properties;
the \emph{application validator} applies any application-specific clock generation constraints.
\sys offers flexibility: an application can choose the appropriate validator instances based on its requirements.

\sys also provides \emph{configurability}.
Different implementations exist for each validator instance:
A validator can be implemented using a group of servers and their signatures, or a single server with trusted execution environment.
To enable this configurability, we further split the abstraction into a \emph{validator frontend} and a \emph{validator backend}.
The frontend defines the \emph{semantics} of a validator instance.
Semantics of a validator defines the verification logic, \ie, what $\var{req}$ and $\var{aux}$ combinations are valid, as discussed above.
They, however, do not define the execution of the verification procedure, which is the responsibility of the validator backend.
Each backend type implements a \emph{security protocol} that generates and verifies \emph{proofs} for the validator frontend logic.
The protocol can use a quorum-certificate-based protocol (\autoref{sec:design:validator-qc}) or leverage trusted execution environment (\autoref{sec:design:validator-tee}), each with different trade-offs in security, performance, and hardware requirements.
The frontends and backends are completely decoupled.
An application can use any backend for its validator instances depending on its specific requirements.

\subsection{Validator Frontend}
\label{sec:design:validator-frontend}

We now elaborate on the frontend logic for the three validator instances.

\subsubsection{Clock Update Validity}
\label{sec:design:validator-update}

The \emph{clock update validator} ensures that a \clk is properly generated using the base clock update rules defined in \autoref{sec:clock:properties}.
For each $\fn{prove}(\var{req}, \var{aux})$ call, $\var{req}$ includes the inputs to a $\fn{Update}(id, c, [c_1, c_2, \cdots])$ invocation, signed by the invoking process, while $\var{aux}$ is empty.
The validator first checks the validity of each input clock by calling $\fn{Verify}()$ on the clock proof $\pi \leftarrow \lbrace\pi_{up}, \pi_{mono}, \pi_{app}\rbrace$, which internally performs $\fn{check}()$ on each validator proof.
Any invalid input clock results in the validator outputting $\bot$.
Next, the validator checks $\var{req}$ is signed by a process with permission on the identifier $id$.
The permission table is either determined statically, or maintained by a global configuration service.
Details of such a service is out of the scope of the paper.
If the permission check passes, the $\var{req}$ is fully validated, and the validator outputs a $\pi_{up}$.
The exact mechanism of generating $\pi_{up}$, as well as verifying $\pi_{up}$, is implemented by the validator backend.

One interesting property of the clock update validator is its \emph{statelessness}.
Besides the permission table (which is either static or global information), the validator does not maintain any state to perform request validation.
Such stateless property has critical implication on the design of certain validator backends, which we will elaborate in \autoref{sec:design:validator-backend}.

\subsubsection{Local Ordering}
\label{sec:design:validator-mono}

The \emph{monotonicity validator} prohibits a process from generating concurrent clocks, or equivalently, enforcing the local ordering property.
The inputs to $\fn{prove}()$ are similar to those of the update validator.
\lijl{Input probably shouldn't be the update function, but just the mapped clock value.}
The validator stores a local table that maintains the highest clock value mapped to each identifier $id$.
It validates the request only if the process has the permission to update $id$, the input clock $c$ is valid and contains a clock value mapped to $id$ which is \emph{no less} than the mapped value in its local table.
If the request is validated, the validator updates its $id$ mapped clock value to an increment of the mapped clock value in $c$, and generates a proof $\pi_{mono}$.
Unlike the update validator, the monotonicity validator is \emph{stateful} since it maintains a clock value table across $\fn{prove}()$ calls.
We will discuss its implications to the validator backend in \autoref{sec:design:validator-backend}.

\subsubsection{Application-Specific Semantics}
\label{sec:design:validator-app}

The \emph{application validator} enforces any additional restrictions the application put on clock updates.
Specifically, $\var{req}$ includes inputs to a $\fn{Update}()$ invocation but without any signature.
$\var{aux}$ contains any application-specific information required to validate the $\fn{Update}()$ request.
The validator runs pre-installed application validation function on $\var{req}$ and $\var{aux}$.
If the function returns true, the validator outputs a proof $\pi_{app}$.
The exact format of $\var{aux}$ and the validation logic are installed on the validators prior to running the application.
We currently restrict the application validator to be \emph{stateless}.
The application-specific validation function therefore is required to be \emph{pure}.
As shown in \autoref{sec:cases}, stateless validation is sufficient for a wide range of distributed applications.

\begin{figure*}
    \centering
    \includegraphics[width=0.8\textwidth]{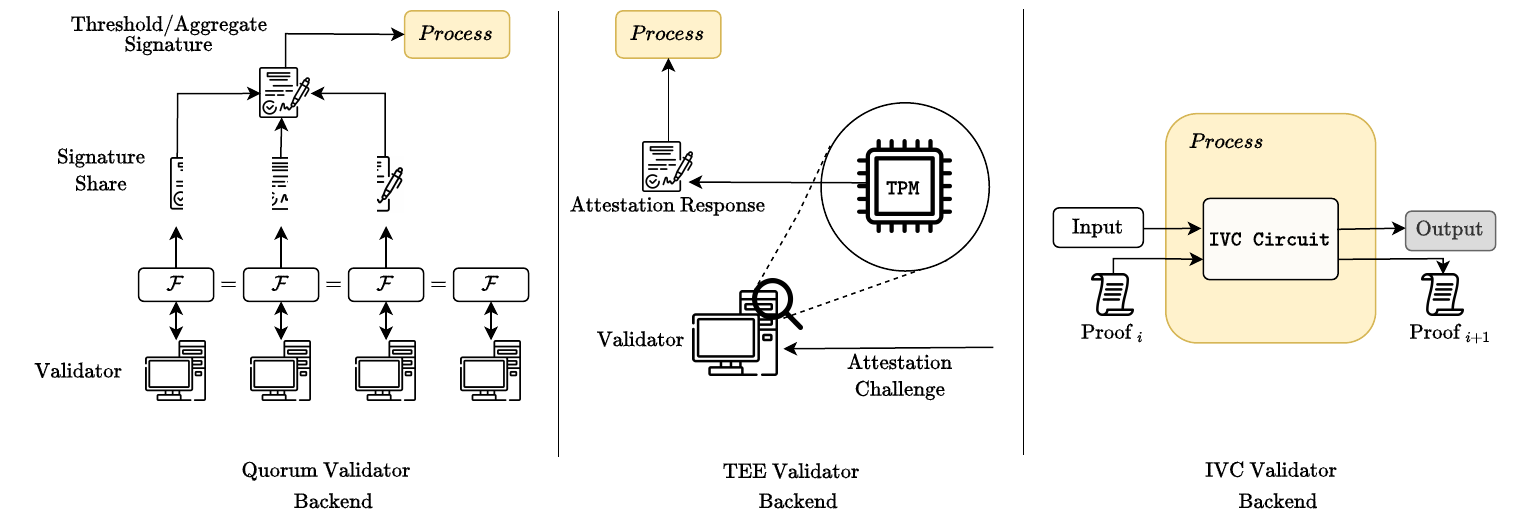}
    \caption{
        Validator backends. \lijl{Add more descriptions to the caption.}
    }
    \label{fig:backend}
\end{figure*}

\subsection{Validator Backend}
\label{sec:design:validator-backend}

As shown in \autoref{fig:backend}, \sys supports multiple forms of backends to generate and verify proofs for each validator instance.
We now elaborate the design of each backend model.

\subsubsection{Quorum Certificate}
\label{sec:design:validator-qc}

The quorum certificate (QC) validator backend consists of a group of servers.
Servers in the group are called \emph{QC validator nodes}.
We assume the group contains $N$ QC validator nodes.
At most $f$ of the nodes exhibits Byzantine failures.
A Byzantine node can behave arbitrarily, but does not have enough computational power to subvert standard cryptographic algorithms.
There exists an external service that manages the group configuration, including nodes joining and leaving the group.
Standard techniques such as proof of stake can be used to tolerate Sybil attacks.
Details of the configuration service is out of the scope of our work.

The group uses a $(t, N)$ threshold signature scheme to generate and verify proofs.
\lijl{Some citations here.}
The choice of $t$ depends on the type of validator frontend logic.
Each validator node applies a distributed key generation scheme to generate its own secrete, distribute shares of the secrete, compute and distribute its part of the combined public key using secrete shares, and generate the final public key by aggregating key parts.
\lijl{Some citations here.}

To get a validator proof, a process sends the type of validator frontend and the input to $\fn{prove}()$, \ie, $\var{req}$ and $\var{aux}$, to $t$ validator nodes.
Each node independently executes the frontend logic.
If the frontend check passes, the node performs the clock update function, creates a partial signature of the output clock using its secret share, and sends the signed clock to the requesting process.
If the process fails to get replies or receives conflicting clocks from the first $t$ nodes, the process contacts other validator nodes from the group.
It waits until it collects $t$ matching clocks with partial signatures from distinct validator nodes.
The process then perform signature aggregation to combine the $t$ partial signatures into a complete signature.
The combined signature on the matching output clock forms the validator proof.
To validate a proof for the $\fn{check}()$ function, any party in the system verifies the combined signature using the aggregated public key from the group.

For stateless validators, the minimum $t$ in $(t, N)$ is $f + 1$.
Since the validator does not need to maintain state across $\fn{prove}()$ invocations, generating a valid proof only requires that \emph{at least one} of the partial signatures with matching output clocks comes from an honest validator.
With $f$ Byzantine validators, setting $t$ to $f + 1$ meets the requirement.
Such $t$ size, however, is not sufficient for stateful validators.
As each $\fn{prove}()$ invocation needs to remember the state of all previous successful $\fn{prove}()$ calls, every two $\fn{prove}()$ invocations need to \emph{intersect} at one or more honest validators.
This requirement can be satisfied by setting $t$ to at least $\lceil \frac{N + f + 1}{2} \rceil$.
\lijl{Is this sufficient?}
\sgd{Probably sufficient as long as a global view on a single point is not required.
For example, it is not sufficient if some application logic specifies whether a clock can be issued is based on \emph{all} clocks that have ever been issued to the id owner, by anyone.
For mono it is probably not the case. It only requires to consider the latest issued clock, not the even earlier ones.}
\lijl{Explain why we don't need consensus for this.}

\lijl{Discuss the complexity when group membership changes.
Need to rerun the key generation scheme.
The history of aggregated public keys must be recorded.
Since the frequency is low, can use consensus.}

\subsubsection{Trusted Hardware}
\label{sec:design:validator-tee}

A trusted execution environment (TEE) validator backend runs on a \emph{single} validator node with access to TEE hardware.
The backend works on any existing implementation of TEE, including ARM TrustZone~\cite{trustzone}, Intel SGX~\cite{sgx}, and AWS Nitro Enclaves~\cite{nitro}.
In this section, we provide a general description of our TEE validator backend design that applies to different TEE variants.
A concrete implementation using Nitro Enclaves is illustrated in \autoref{sec:impl}.

The code implementing the validator frontend logic is released to the public.
Any third party can independently verify the cryptographic hash of the code.
A validator node with access to TEE hardware creates and launches an enclave image that runs the frontend code.
To process a $\fn{prove}(\var{req}, \var{aux})$ call, the validator node passes the call inputs to the running enclave.
The exact information passing mechanism depends on the TEE implementation.
The enclave then performs the frontend logic on the inputs.
If the operation is properly validated, the enclave computes the output clock, and requests an attestation document from the TEE hardware.
The attestation request includes the output clock as user data.
The resulting attestation document serves as the validator proof of the output clock.

To verify a proof in the $\fn{check}()$ function, any participant in the system first validates the authenticity of the attestation document.
Trust of the attestation certificate authority is established through external mechanisms (\eg, AWS root certificate).
Once the document is authenticated, the verifier checks that the enclave image in the document matches the hash of the publicly available frontend code.
It then validates that the clock matches the user data field in the document.
The verifier outputs $\var{true}$ only if these tests all pass.

Unlike the QC backend, \emph{any single} TEE validator can generate proofs for stateless validator frontends.
A valid attestation document guarantees that the clock is verified by a correct program that implements the requested frontend logic.
The trusted computing base (TCB), in this case, is reduced to the TEE hardware and its root of trust (\eg, Intel, ARM, or AWS).
Implementing stateful frontends, however, still requires a quorum out of $N$ TEE validators, as the latest state might be lost when using a single validator.
But since no TEE validator can produce erroneous outputs, only a simple majority of matching validator outputs, \ie, $\lceil \frac{N + 1}{2} \rceil$, is required to generate a proof for stateful frontends.

\subsubsection{Verifiable Computation}
\label{sec:design:validator-ivc}

\begin{figure}
    \centering
    \includegraphics[width=0.5\textwidth]{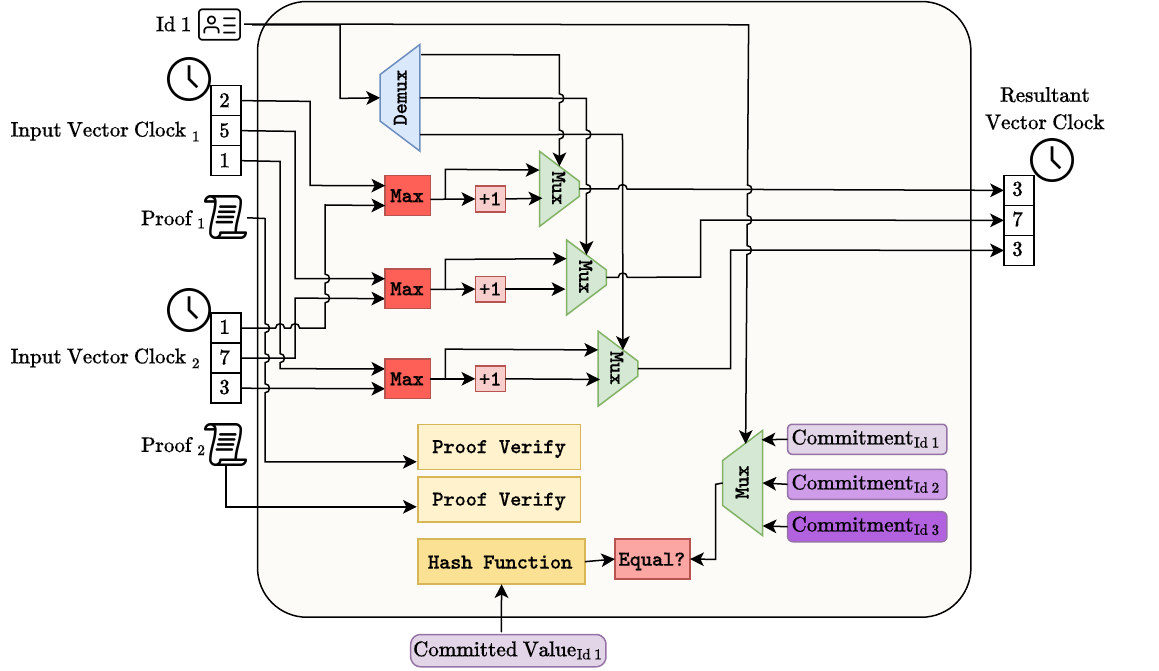}
    \caption{
        Design of verifiable computation based validator backend.
    }
    \label{fig:ivc}
\end{figure}

The VC backend runs the entire frontend logic using verifiable computation (VC).
Our \emph{recursive verifiability} requirement puts restrictions on the type of VC that we can use.
Specifically, the update validator needs to verify multiple input clock proofs;
this requires a proof carrying data (PCD)~\cite{pcd} supported VC technique, such as recursive SNARK~\cite{snarks} and HyperNova~\cite{hypernova}.
For stateless validators, any general VC solution also apply.
However, due to the limitation of VC, stateful validators cannot be implemented using a VC backend.

The frontend logic is first translated into arithmetic circuit.
\autoref{fig:ivc} shows the simplified circuit architecture of a clock update frontend.
In the circuit, a clock $c$ is updated by merging with one other clock $c_1$, and both clocks hold counter values for 3 $id$s.
The circuit takes four witnesses.
The first three are $c$, $c_1$ paired with their proofs, and the updated $id$.
These witnesses match the arguments of $\fn{Update}$, and will be used during verification as well.
The circuit additionally takes a committed value for the updated $id$ as witness, to authenticate the processor's identity.
The witnesses are not required during verification, but a proving process can only be done if the processor can provide valid witnesses.

In the circuit, the two clock proofs are verified while the clock values are maximized element wise.
Next, the counter value mapped to $id$ will be incremented using a group of muxs and a demuxs.
The committed value is then verified against the corresponding commitment programmed into circuit statically.
Lastly, a proof is generated along with the output clock value.

The circuit has only one public input, which is the output clock value.
Any participant can validate the clock by verifying the proof using the clock value as public input.
Similar to the trusted hardware backend, a single execution of the VC is sufficient to generate proofs.
\section{Use Cases}
\label{sec:cases}

In this section, we describe two use cases of \clk: a mutual exclusion service and a weakly consistent data store.
Both systems are evaluated in \autoref{sec:eval}.

\subsection{Mutual Exclusion}
\label{sec:cases:mutex}
\sgd{Add locking case}

Our first use case is inspired by the mutual exclusion algorithm in Lamport's paper~\cite{lamport-clock}.
The system consists of $N$ processes and a shared resource.
Each process may request to access the shared resource.
A mutual exclusion distributed algorithm needs to satisfy the following properties:

\begin{itemize}
    \item \emph{Mutual exclusion}: Access to $r$ is granted to \emph{at most one} process at any time.
    \item \emph{Ordered access}: For any two resource access requests $r_1$, $r_2$ and their respective logical clocks $c_1$ and $c_2$, if $c_1 \prec c_2$, then $r_1$ is granted before $r_2$.
    If $c_1 \parallel c_2$, the two requests can be granted in any order.
    \item \emph{Starvation free}: If every process eventually releases the resource after it is granted access, all access requests are eventually granted.
\end{itemize}

\lijl{Starvation free is only possible if network is partially synchronous, and that all nodes respond to messages. For the second point, can add incentive-based solution in the discussion.}

\lijl{The ordered access property needs to be refined.
A Byzantine node can always use older logical clocks or concurrent clocks to gain unfair advantage.
Need to include some fairness property.}
\sgd{Resource provider can always detect for it through examining the proof right?}

Given that Byzantine processes can ignore the mutual exclusion protocol and access the shared resource at any time, we require the process accessing the resource to present a \emph{proof} to demonstrate its eligibility.
Owner of the resource verifies the proof before granting it to the process, although details of this interaction is out of the scope our protocol.

There are five message types in our protocol: \mtype{Request} for requesting the shared resource, \mtype{Reply} for responding to \mtype{Request}, \mtype{Release} for releasing the shared resource, \mtype{Query} for checking other processes, and \mtype{Ack} for responding to \mtype{Query}.
Each message is tagged with the identifier of the sending process and a clock;
a \mtype{Reply} message additionally contains a list of clocks, with the list size equals the number of processes.
Each process locally maintains a local clock and a queue of \mtype{Request}{}s it has received (including from itself).
It also stores the most recent clock it has received from each other process.

We define a total order $\lessdot$ on the set of clocks.
$\lessdot$ strictly contains the partially ordered set $\prec$, i.e., if $c_1 \prec c_2$ then $c_1 \lessdot c_2$.
For concurrent clocks, $\lessdot$ defines some arbitrary ordering rules.
Details of the rules do not impact the core properties of the protocol.
\lijl{Does it impact fairness?}

When a process sends any message, it attaches its local clock $c$ into the message.
Upon receiving any message with a clock $c_m$, the process first verifies $c_m$.
It then updates its local clock $c$ to the result of $\fn{Update}(id, c, [c_m])$.
\lijl{Do we care about duplicated messages? Safety-wise it should still be fine.}

To request access to the shared resource, a process
broadcasts a \mtype{Request} message and adds the message to its local queue.
When a process $p$ receives a \mtype{Request} $R$ with a valid clock $c_r$, it adds $R$ to the queue.
If the requesting process already had a \mtype{Request} in the queue, $p$ ignores $R$.
For each other process $q$, $p$ then waits until either it has a \mtype{Request} from $q$ in the queue, or its most recent known clock from $q$ happened after $c_r$ (by sending a \mtype{Query} message to $q$ and waits for the \mtype{Ack}).
$p$ then responds with a \mtype{Reply} message.
The \mtype{Reply} message contains the clock of each \mtype{Request} ordered before (by $\lessdot$) $c_r$ in the queue.

A process acquires the shared resource when the following conditions are met.
1) Its own \mtype{Request} has the smallest clock, ordered by $\lessdot$, among all \mtype{Request}{}s in the queue.
2) It has received a message with a clock ordered after (by $\prec$) its \mtype{Request} clock from every process.

To release the resource, a process removes its own \mtype{Request}{}s in the queue, and broadcasts a \mtype{Release} message.
Once a process receives a \mtype{Release}, it verifies that the \mtype{Release} is ordered after its pending \mtype{Request} (if any) and any \mtype{Request}{}s from the releasing process in its local queue.
If the conditions are met, the process removes all \mtype{Request}{}s from the releasing process in the queue.
Note that \mtype{Request}{}s and \mtype{Release}{}s from a single process is always totally ordered.
After a process $p$ receives a \mtype{Release} message from $q$, $p$ can ignore further \mtype{Request}{}s from $q$ that is \emph{not} ordered after the \mtype{Release}.

To access the shared resource, a process presents an \emph{acquisition proof} containing its own \mtype{Request}, and either a \mtype{Reply} or a \mtype{Release} from every other process.
The resource owner validates the proof by checking that all \mtype{Release} messages have clocks ordered after (by $\lessdot$) the \mtype{Request}.
Moreover, for each clock $c$ in each \mtype{Reply}'s clock list, the corresponding process has a \mtype{Release} message in the proof.

\subsection{Causally Consistent Data Store}
\label{sec:cases:cstore}

Logical clocks have been used to build data stores that provide causal consistency~\cite{cops}.
Here, we leverage \clk to construct a similar data store that tolerates Byzantine processes.

The system consists of $N$ servers, collectively implementing a distributed key-value service.
Each server maintains the entire key-value mappings, \ie, the key-value store is fully replicated on $N$ servers.
\lijl{Can add sharding support.}
The service exposes $\fn{Get}(k)$ and $\fn{Put}(k, v)$ API calls to clients.
It provides causal consistency.
Each $\fn{Put}$ implicitly contains all previously retrieved object versions from $\fn{Get}()$ calls within the client session.
These object versions are \emph{causal dependencies} of the new object version created by $\fn{Put}$.
All $\fn{Get}()$ respect causal ordering: if a client observes an object, it is guaranteed to later observe all its causally dependent object versions (or more update-to-date versions).

\sgd{Updated: hide clocks from user interface; transparently guarantee safety}
\sgd{Should we a. more formally define \emph{observable} or b. completely bypass the concept of observable}

\sgd{Added.}

Both of the APIs implicitly refers to and maintains a \emph{session context} on client side, which is used to determine a range of dependent operations of every operation.
In a session, operations always causally depend and only depend on all operations that are finished prior to their invocations.
For simplicity, all $\fn{Put}$ of a session happen sequentially.

The key space is divided into $N$ subspace, each assigned to a server.
Clients send $\fn{Put}$ requests to the server responsible for the key.
This key partitioning design is to eliminate conflict resolution.
\lijl{Byzantine servers can still create conflicts. Other approaches are required to fix Byzantine conflicts, such as slashing.}
Partitioning metadata is managed by a BFT consensus protocol.
However, $\fn{Get}$ requests can be sent to any of the $N$ servers.

\sgd{Updated.}

Each client maintains a \emph{dependency clock} for a session.
\sgd{Not talking about the details of dependency clock is an incomplete clock while versioned dependency clock is a complete clock yet.
We need this distinction because client can only do MERGE on dependency clock while server can do MERGE\_AND\_INC on versioned dependency clock.
Complete these details after clock section update.}
The clock records the aggregated causal dependency of the session, and is initialized with $\fn{Init}()$ at the beginning of the session.
Unlike other clocks where the clock value is considered opaque, individual counter of a dependency clock may be individually checked during verification.
Every key corresponds to an $id$, and we use $c[id]$ to refer the counter value of $id$ in clock $c$.
In a dependency clock $c$, $c[id] = ver$ means the operation depends on the version $ver$ of the key $id$.
A client can check whether the result of a $\fn{Get}(k)$ fulfills the causality requirement by checking whether the returned value has a version that is not older than $c[id]$ of its dependency clock, where $id$ corresponds to $k$.

\sgd{Added.}
Each server maintains a \emph{versioned dependency clock} associated with the current value of each key.
The versioned dependency clock has the same definition as the dependency clock, except that for the $id$ that corresponds to the mapping key, the counter $c[id]$ stores the version number of the key.

It's easy to see that
\begin{itemize}
    \item Given a dependency clock $c$, if for every $id$, the versioned dependency clock of the corresponded key shows that the version number is not less than $c[id]$, then the server is \emph{up-to-date} regarding $c$, that is, everything from now on happens on this server is logically ordered after all the dependent operations involved in $c$.
    This is the prerequisite of serving requests with dependency clock $c$.
    \item If $\fn{Compare}$ shows that a versioned dependency clock $c$ is greater than a dependency clock $c'$, than the value clocked $c$ is considered to be logically ordered after all the dependent operations involved in $c'$.
    This is the prerequisite of replying a $\fn{Put}$ request with versioned dependency clock $c$.
\end{itemize}

\sgd{Added.}
For a $\fn{Put}$ operation, the client attaches its current dependency clock in the $\fn{Put}$ request.
Server should reply both $\fn{Get}$ and $\fn{Put}$ requests with a versioned dependency clock.
For $\fn{Get}$, it is the clock of the returned value $v$; for $\fn{Put}$, it is the newly created clock for the requested $v$.
Client first verifies the returned clock, then checks for causal dependencies according to its local dependency clock $c$.
For $\fn{Get}(k)$ request, the returned value must have a version number that is not less than $c[id]$ where $id$ corresponds to $k$.
For $\fn{Put}$ request, the returned versioned dependency clock must be greater than the local dependency clock ($AF$ in $\fn{Compare}()$).
The server reply is only accepted when both the clock is valid and the causal dependencies are respected.
Client then merges the returned clock into its local dependency clock for the following operations.

\sgd{Updated}
When server receives a client request, it first checks whether its local state is up-to-date according to the definition above.
The request is rejected if the check fails.
\sgd{This is only the case when the session is served by single server. Need more discussion on this.}
\sgd{Also "as long as" implies necessary but not sufficient; the "external causality" that happens to be able to be satisfied is also covered here.}
Since causal dependencies are always satisfied, server can immediately reply to the requests.
For $\fn{Get}$ requests, server simply returns its stored value and its versioned dependency clock.
For $\fn{Put}$ requests, server creates a new clock by applying $\fn{Update}()$, stores the created clock with the requested value, and returns the clock to client.
\sgd{Clock operation name TBD.}

\sgd{The client signature on the dependency clock is used as (part of) the input of the "application blackbox" of $\fn{MergeInc}$.
The application blackbox is not mentioned at all in current text. Well, for good maybe.}

\sgd{Updated.}
Once a server installs a new versioned dependency clock and a value for a key, it asynchronously propagates this key-value entry to all other servers.
Upon receiving the entry, a server need to ensure causal consistency.
It does so by first examines whether its local state is up-to-date based on the versioned dependency clock.
If not, it puts the propagated entry into pending state.
Otherwise, it can install the key, and recursively reexamine and commits the pending propagated entries.

%
\section{Implementation}
\label{sec:impl}

We have implemented the three validator backends as described in \autoref{sec:design:validator-backend}.
The QC backend uses the schnorrkel Rust library~\cite{schnorrkel} which implements Schnorr signature on Ristretto compressed Ed25519 points.
The library provides half-aggregated (\ie prepared) batch signature verification functionality~\cite{chalkias2021non}.
It enables cryptographic overhead sublinear to the number of signatures in a QC.

The trusted hardware backend is implemented on AWS Nitro Enclaves~\cite{nitro}.
Nitro Enclaves use the same Nitro Hypervisor technology
to isolate the vCPUs and memory for an enclave from its parent instance.
Enclaves establish secure local socket connectivity (using Linux VSOCK) with their parent instances.
Nitro Enclaves also supports attestation: Using the Nitro Enclaves SDK, an enclave can request a signed attestation document from the Nitro Hypervisor that includes its unique measurements.
The validator requests its own measurements, \ie platform configuration registers (PCRs), on startup.
A clock update is only considered to be valid if all the included clock are attested with the same PCRs.
To generate a proof, the validator requests an attestation document from the hypervisor for the new clock value.
The validator program is built into an enclave image file, and is deployed with Docker using Nitro Enclaves CLI.

The verifiable computation backend is implemented using Plonky2~\cite{plonky2}.
Leveraging the universal verification circuit present in Plonky2, we build a clock circuit that verifies two input proofs which are proved by itself.
\sgd{maybe add a gate breakdown here if has space}
We have implemented the update validator logic in all backends;
monotonic and application validators are left as future work.

We implement the mutual exclusion use case atop a causal network protocol.
The causal network acts as a middlebox:
For egress messages, it assigns the process current logical clock to the messages;
for ingress messages, it updates the local clock using the clock update function.
The application then only considers message send and receive events with causal ordering.
The acquisition proof consists of configured number of Ed25519 signatures which uses the same schnorrkel library as QC validator.

We implement COPS following the original protocol~\cite{cops}.
Our implementation replicates asynchronously to remote clusters, while each local cluster works as a replicated state machine using PBFT~\cite{pbft}.
The COPS-GT and COPS-CD variants are omitted from our implementation.

The entire codebase is implemented with 6.5k lines of Rust code.
The mutex and COPS use cases take 595 and 585 lines of code respectively.
The QC and Nitro Enclave backends take 716 lines of code in total.
The plonky2 backend takes 408 lines of code.
The comparison PBFT implementation takes 1274 lines of code.
\section{Evaluation}
\label{sec:eval}

Implementation details of \sys can be found in \autoref{sec:impl}.
We performed end-to-end evaluations on the two use cases (\autoref{sec:cases}) and micro-benchmarks on validator backends (\autoref{sec:design:validator}).
We set up realistic global P2P network for performing our end-to-end evaluations.
Identical instances are spawned across 5 AWS regions (ap-east-1, us-west-1, eu-central-1, sa-east-1 and af-south-1).
The instances are interconnected through public Internet even in the same region.
We compare \sys with our own PBFT~\cite{pbft} implementation in end-to-end evaluations.
PBFT is used as a Byzantine fault-tolerant ordering service that provides \emph{total ordering}.

In end-to-end evaluations, we evaluate \sys using both QC and TEE validator backends.
\sgd{That only verify the basic properties.}
For QC, we spawn two c5a.16xlarge instances in each region to simulate a validator that has sufficient computation and network resources.
We set the quorum size to two
to tolerate one faulty quorum member.

\subsection{Mutex Latency}

\begin{figure}
    \hspace*{0.23in}
    \includegraphics[width=0.23\textwidth]{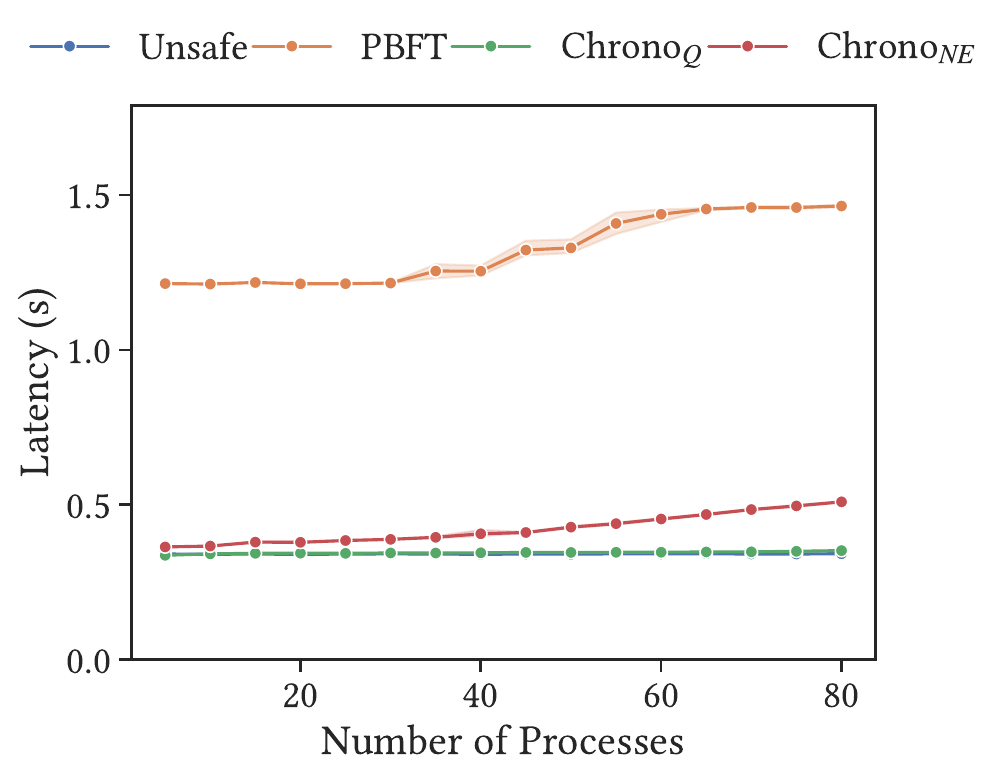}
    \includegraphics[width=0.2\textwidth]{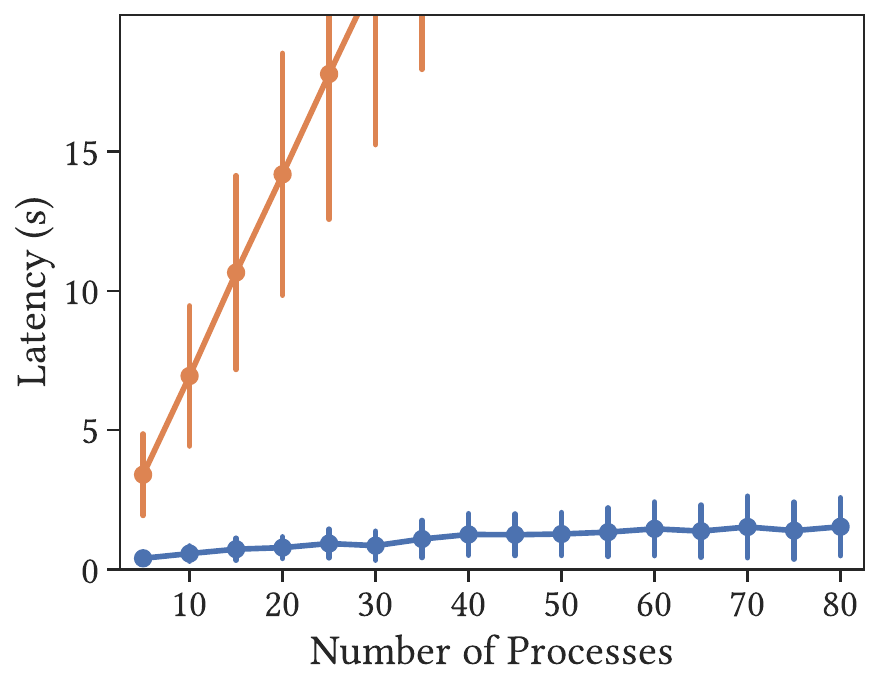}
    \caption{
        Resource acquisition latencies in the mutex use case.
    }
    \label{fig:mutex}
\end{figure}

We first study the latency of the mutex protocol (\autoref{sec:cases:mutex}) with varying number of processes.
We spawn a c5a.xlarge instance with four vCPUs for each process, and evaluate the QC and Nitro Enclaves backends.
Two of the four vCPUs are allocated to Nitro Enclaves.
We additionally spawn two c5a.8xlarge instances with 32 vCPUs per region for quorum nodes.

\autoref{fig:mutex} compares \sys with two other systems: the baseline original protocol~\cite{lamport-clock} which tolerates no Byzantine failures, and a protocol where processes communicate through PBFT-based broadcast so that messages are received in the same total order on all processes.
We evaluated two scenarios: only one of the processes keeps requesting the resource, and all processes request simultaneously.
In both cases, the process immediately release the resource after it is successfully granted the lock.

In the single process request experiment, all four systems maintain near constant latency.
\sys with QC backend achieves a latency that is virtually identical to the baseline, and the Nitro Enclaves backend only induces marginal additional latency.
In contrast, PBFT has significantly higher latency due to its sequential nature.
Although it can merge \mtype{RequestOk{}} messages into one batch to avoid growing latencies, every point-to-point message requiring two all-to-all PBFT message rounds still leads to amplified latency.

When all processes are contending for resources, request latency can be divided into two phases: before and after the first process is granted with the resource.
The former phase is complete concurrent.
A system with better concurrency can reduce the average latency of this phase.
The second phase is complete sequential.
A system with shorter step latency can benefit from this phase, resulting in a smaller latency standard deviation.
As shown in the \autoref{fig:mutex}, PBFT performs subpar in both phases.
The two variants of \sys, while achieves similar standard deviation as the baseline, experiences significant longer average latency.
This is due to the quadratic per-process workload required for generating acquisition proofs.
We argue that this cost is justifiable for tolerating Byzantine behaviors.

\subsection{Causal Consistency Data Store}

\begin{figure}
    \hspace*{-0.23in}
    \centering
    \includegraphics[width=0.25\textwidth]{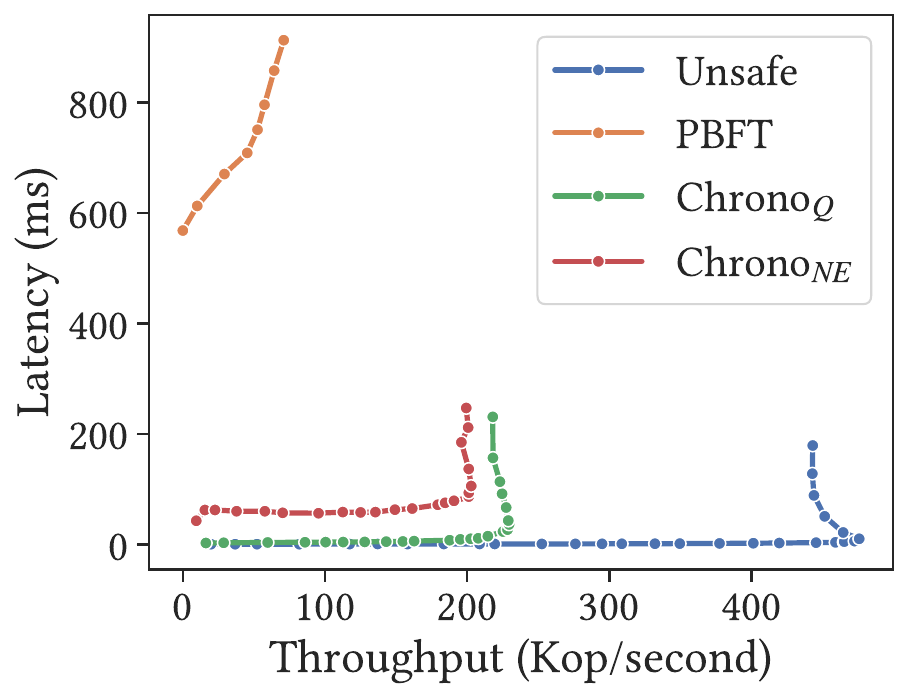}
    \includegraphics[width=0.25\textwidth]{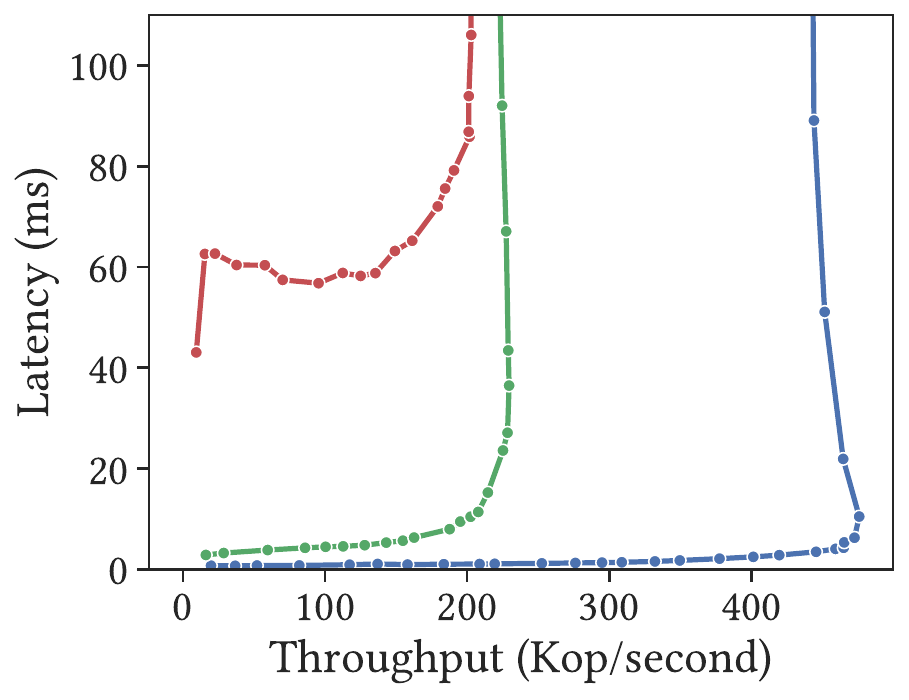}
    \caption{The throughput-latency relation of causal consistency data store under different workload.
    The same results are plotted with different latency range.}
    \label{fig:throughput-latency}
\end{figure}

Next, we evaluate the throughput of our logical-clock-based causally consistent data store (\autoref{sec:cases:cstore}).
We deploy five causally consistent replicas, one in each region, on c5a.8xlarge instances with 32 vCPUs and 10 Gbps network.
We also evaluate the QC and Nitro Enclaves backends in this evaluation.
On each replica, 20 vCPUs are preserved for enclaves throughout the evaluation.
We set up two quorum instances per region with the same spec as the mutex evaluation.

The system described in the original paper~\cite{cops} is used as the unsafe baseline.
We also directly compare to a geo-replicated PBFT deployment.
The PBFT protocol replicates every \fn{Get} and \fn{Put} to all regions, and the total order of all requests fulfill the causal consistency requirement.
Notice that the security model of PBFT is difference from \sys.
PBFT can only tolerate one faulty replica among the total five replicas, while \sys can still ensure causality even if all five replicas are faulty.
On the other hand, PBFT can ensure stronger availability, while \sys can only tolerate causality related faulty behaviors.

We use a mix of YCSB workload A (50:50 read/write) and C (read-only) to control the overall read/write ratio of the workload.
Workload A is modified to uniformly distribute access across local keys following the original work~\cite{cops}.
We use the workload with 1\% of write requests by default.

\autoref{fig:throughput-latency} shows the 99.9th latency of all evaluated systems.
PBFT is not comparable to the other systems in both throughput and latency.
For latency, each request must reach every region before the reply can be generated, while in all other systems only reply from the local replica is needed.
For throughput, although our PBFT implementation already amortizes the replication overhead using batch size of 100, the protocol complexity still exposes significantly higher overhead than weaker consistency models.
This is particularly true for the \fn{Get} requests, which only require simple hashmap lookups in other systems.

We then zoom into the results of the unsafe baseline and the two \sys variants.
Although the Nitro Enclaves backend incurs higher minimal latency due to hardware factors, the system eventually saturates at around 200Kop/s.
This performance is close to that of the QC backend, since the two variants share the same COPS implementation, so the critical protocol path is identical.
With sufficient computational resources on the backends, we should expect other backend variants to achieve similar throughput.

\begin{figure}
    \hspace*{-0.3in}
    \centering
    \minipage[t]{0.25\textwidth}
        \includegraphics[width=\linewidth]{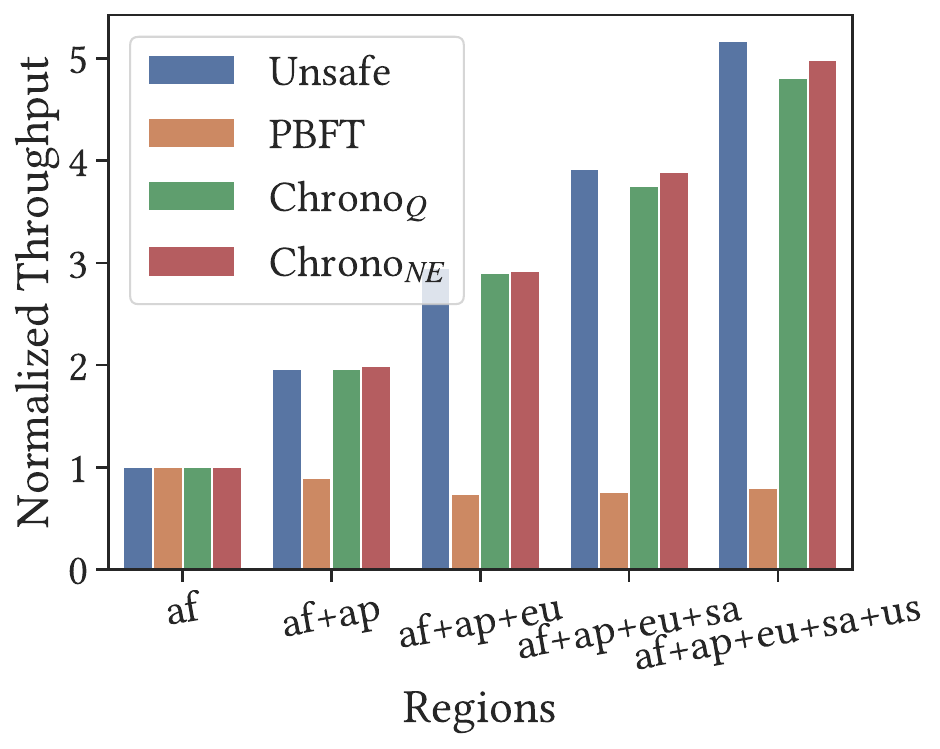}
        \caption{Normalized maximum throughput when deploying across difference number of regions.
        }
        \label{fig:region}
    \endminipage\hspace{0.05in}
    \minipage[t]{0.25\textwidth}
    \raisebox{0.08in}{
        \includegraphics[width=\linewidth]{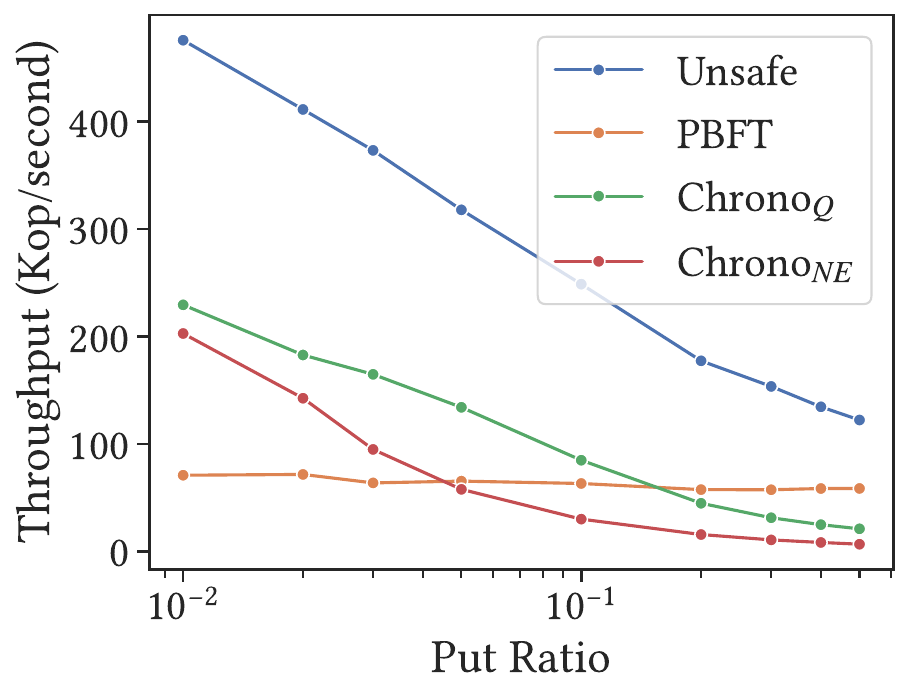}
    }
        \caption{Maximum throughput under workloads of different read/write ratios.
        }
        \label{fig:ratio}
    \endminipage

\end{figure}

\autoref{fig:region} shows the maximum throughput when deploying all systems with increasing number of regions.
The results are normalized to the maximum throughput of the same system when deployed in only af-south-1 region.
As expected, both the unsafe baseline and the variants of \sys perfectly scale against the number of regions, because all cross region communication are asynchronous and not on the critical path.
On the contrary, PBFT does not scale.
Its performance even drops slightly when going beyond a single region, which is attributed to the overhead introduced by imperfect network conditions (\eg, reordering).

The \autoref{fig:ratio} shows the throughput of the systems under workloads with different read/write ratios.
Both \sys and unsafe baseline encounter performance degradation when there are more write requests, due to the more complex code path for handling \fn{Put}.
The QC variant can maintain higher throughput than PBFT with less than 20\% \fn{Put} requests in the workload, while Nitro Enclaves variant drops below PBFT with 5\% \fn{Put}, as the computation inside enclaves becomes the system bottleneck.
The bottleneck, however, can be eliminated by provisioning more resources to the Nitro Enclaves backend.

\subsection{Micro-benchmarks}

\begin{figure}
    \hspace*{-0.23in}
    \centering
    \includegraphics[width=0.25\textwidth]{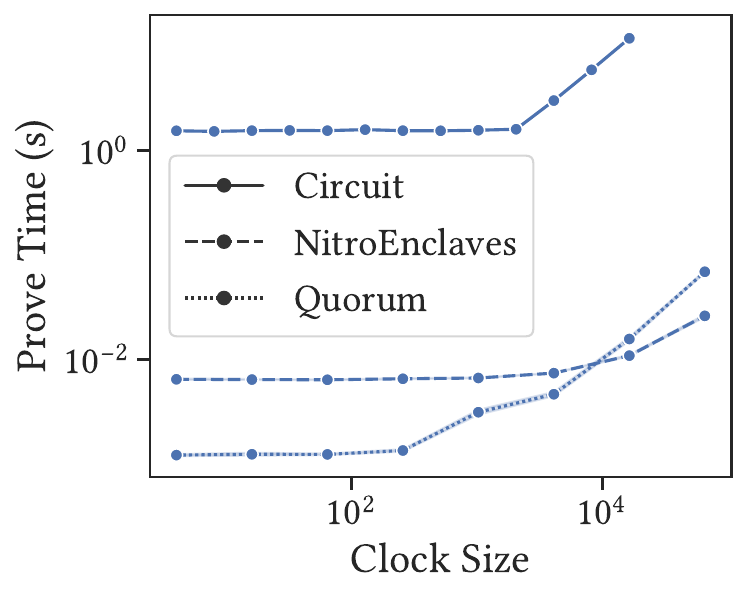}
    \includegraphics[width=0.25\textwidth]{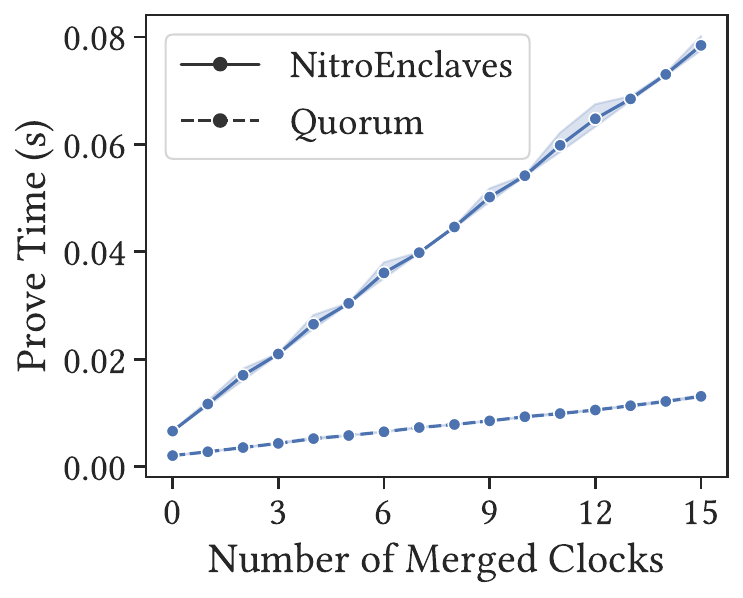}
    \caption{Proving time with varying clock size and number of clock merged into the updated clock.
    The circuit variant is measured with 64 CPUs.
    }
    \label{fig:clock-prove}
\end{figure}

In our first set of micro-benchmarks, we explore the factors that affect backend performance with fixed computational resource.
The first factor is the clock size, which means how many $id$s the clock is keeping track of the causal dependencies.
For circuit backend, the maximum clock size must be hardcoded ahead of time in order to synthesis arithmetic circuit accordingly, and the backend will always keep track of all $id$s no matter they are occupied or not.
So in this evaluation, we build circuits for different configured maximum clock sizes and perform update operation on them.
As for quorum and Nitro Enclaves backends, the map-based clock data structure allows dynamically adding more $id$s, so we preload the clocks with the expected number of $id$s before measurement.

All three variants maintain constant proving time before clock size grows over certain threshold.
Before start to grow, circuit variant has second level latency, while both quorum and Nitro Enclaves variants have millisecond level latency.
Quorum variant's latency increases faster than Nitro Enclaves variant due to lower throughput transmitting to remote quorum node compare to Nitro Enclaves' local Vsock connection even in the absence of concurrency.

The second studied factor is the number of merged clocks inputted into \fn{Update} operation.
Circuit cannot take varying number of merged clocks as input due the limitation of verifiable computation must be finished in fixed number of steps, so we focus on the other two variants in this micro-benchmark.
The results from both variants show linearly growing latency regarding the number of merged clocks.

\begin{figure}
    \hspace*{-0.16in}
    \centering
    \includegraphics[width=0.24\textwidth]{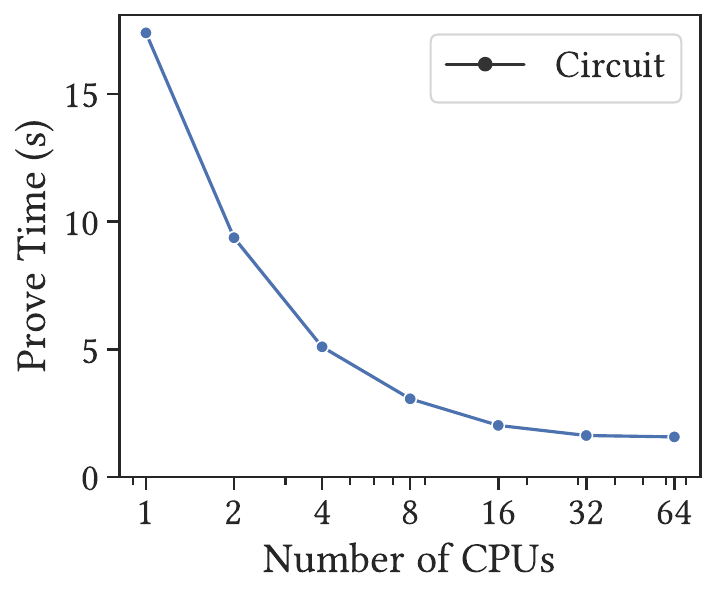}
    \includegraphics[width=0.25\textwidth]{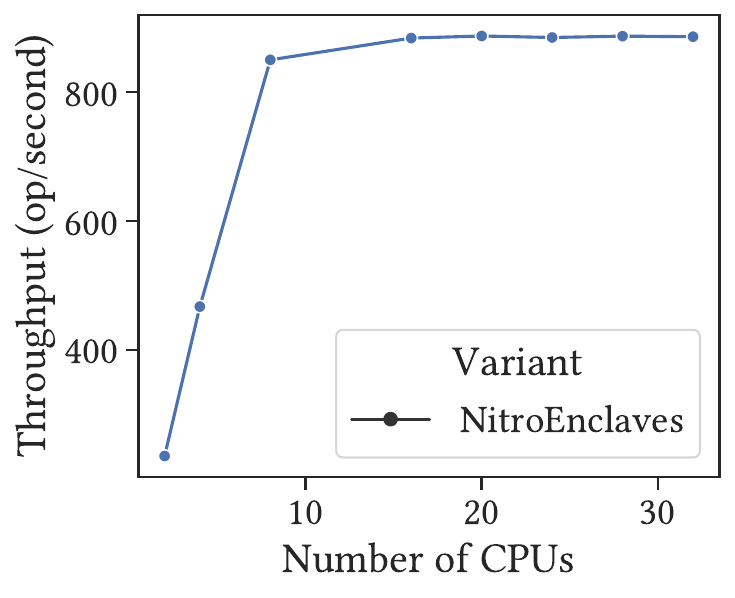}
    \caption{Proving performance with varying computational resource.
    }
    \label{fig:clock-cpu}
\end{figure}

In the next set of micro-benchmarks we study the backend performance under varying computational resource.
As shown in \autoref{fig:clock-cpu}, both circuit and Nitro Enclaves backends scales with the number of CPUs.
However, they cannot scale perfectly as CPU number increases.
For circuit backend, the sequential computation will eventually dominate, and for Nitro Enclaves, the backend eventually bottlenecks on the secure socket between host and VM.
Both backend can scale well up to 20-30 CPUs.

\begin{figure}
    \centering
    \includegraphics[width=0.25\textwidth]{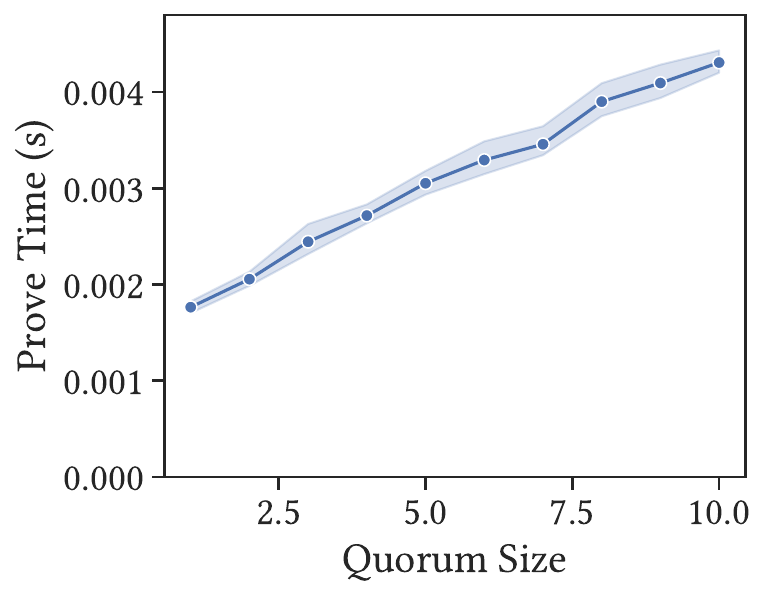}
    \caption{Proving time with different quorum size.
    }
    \label{fig:clock-quorum}
\end{figure}

Lastly we also perform a dedicated micro-benchmark on the most performant quorum variant, studying how the performance will be affected when tolerating more and more faulty quorum members.
The result shows a linear growth, mostly due to the longer verification time of the quorum certificates that includes more signatures, on both quorum and user side.
\section{Conclusion}
\label{sec:concl}

In this work, we introduce a new logical clock system, \sys.
\sys addresses key limitations of prior approaches that fail to work in networks with Byzantine participants.
It includes a novel logical clock structure, \clk, that builds atop a \emph{validator} abstraction.
The abstraction defines the semantics of correct logical clock updates.
Various backend implementations, each with different security-performance trade-offs, then enforce these validator semantics.
To demonstrate the power of \sys, we developed two concrete applications: a mutual exclusion protocol and a causally consistent data store.
Evaluation results showed that \sys is an attractive solution for building decentralized applications and systems in an open P2P network.
 
\bibliography{paper}

\end{document}